\documentclass[%
12pt,
eprint,
groupedaddress,
 amsmath,amssymb,
aip,
 onecolumn
]{revtex4-1}

\usepackage[utf8]{inputenc}
\usepackage[english]{babel}
\usepackage{amsthm,textcomp}
\usepackage{amsmath}
\usepackage{amssymb}
\usepackage{physics}
\usepackage{bbm}
\usepackage{stackengine,graphicx}   
\usepackage{pgf}
\usepackage{subcaption}
\usepackage{qcircuit}
\usepackage{hyperref}
\usepackage[export]{adjustbox}
\usepackage{tikz}

\usepackage{dashrule} 

\usepackage{color,soul}

\usepackage{todonotes}

\renewcommand{\S}{Section}

\newcommand{\rmi}{\mathrm{i}} 

\DeclareMathOperator{\Var}{Var}

\tikzset{blank/.style={rectangle,inner sep=0pt,draw=none,fill=none,minimum
    size=0pt} }

\begin{document}

\title{On barren plateaus and cost function locality in variational quantum algorithms}
\author{A.\,V.~Uvarov}
\email{alexey.uvarov@skoltech.ru}
\author{J.\,D.~Biamonte }
\email{j.biamonte@skoltech.ru}
\thanks{webpage: https://quantum.skoltech.ru}
\affiliation{Skolkovo Institute of Science and Technology, 
3 Nobel Street, Moscow 143026, Russian Federation}

\date{September 2020}

\newtheorem{lemma}{Lemma}
\newtheorem{theorem}{Theorem}
\newtheorem{proposition}{Proposition}
\newtheorem{corollary}{Corollary}

\theoremstyle{definition}
\newtheorem{definition}{Definition}
\newtheorem{example}{Example}

\theoremstyle{remark}
\newtheorem*{remark}{Remark}

\newcommand{\mc}[1]{\mathcal{#1}}

\sloppy

\begin{abstract}
    Variational quantum algorithms rely on gradient based optimization to iteratively minimize a cost function evaluated by measuring output(s) of a quantum processor. 
    A barren plateau is the phenomenon of exponentially vanishing gradients in sufficiently expressive parametrized quantum circuits.
    It has been established that the onset of a barren plateau regime depends on the cost function, although the particular behavior has been demonstrated only for certain classes of cost functions.
    Here we derive a lower bound on the variance of the gradient, which depends mainly on the width of the circuit causal cone of each term in the Pauli decomposition of the cost function.
    Our result further clarifies the conditions under which barren plateaus can occur.
\end{abstract}

\maketitle 

\section{Introduction}


While current quantum hardware maintains certain limits on its capabilities, modern quantum algorithms are designed to circumvent these limitations. A prominent family of such device tailored algorithms---called hybrid
quantum-classical or variational quantum algorithms \cite{mcclean_theory_2016}---uses the quantum device in tandem with a classical computer: typically, a cost function is evaluated by preparing and then measuring a quantum state.  The parameters of the prepared state are iteratively updated to minimize the cost function. Examples of such algorithms include variational quantum eigensolver (VQE) \cite{peruzzo_variational_2014, kandala_hardware-efficient_2017, barkoutsos_quantum_2018, cade_strategies_2019}, quantum approximate optimization algorithm and variational quantum search (QAOA) \cite{farhi_quantum_2014,willsch_benchmarking_2019,morales_variational_2018, Akshay_Philathong_Morales_Biamonte_2020}, quantum autoencoders \cite{romero_quantum_2017}, and training of quantum neural networks \cite{havlicek_supervised_2019,schuld_circuit-centric_2018,huggins_towards_2019,uvarov_machine_2019}.  

Several families of variational circuits have been studied and trained to minimize cost functions.  These include the hardware efficient ansatz (HEA) \cite{kandala_hardware-efficient_2017}, the checkerboard or brick-layer tensor network (CBA) \cite{brandao_local_2016, cerezo_cost-function-dependent_2020, Nakaji_Yamamoto_2020, Li_Chen_Fisher_2019}, the unitary coupled cluster truncated at various orders (UCC) \cite{wecker_progress_2015, romero_strategies_2017, taube_new_2006}, tree tensor networks (TTN) as well as the alternating operator ansatz found in QAOA.  Most of these gate sequences can in principle be tuned to mimic general quantum circuits.  Interestingly, long QAOA sequences can indeed emulate general quantum circuits \cite{lloyd_quantum_2018,morales_universality_2019}. The variational approach has further restrictions in that the output of the quantum computer must be evaluated with respect to an objective function ascertained by local measurements. This approach to quantum computation represents a universal model and is hence as powerful as any quantum computer would be \cite{biamonte_universal_2019}. In practice however, the computational capacity of the variational model is not fully understood.  The available gate sequences are short and objective function minimization can require significant resources in the classical optimization step and might not even be possible, due to e.g.~reachability deficits \cite{Akshay_Philathong_Morales_Biamonte_2020}.  


The classical optimization step, integral to these algorithms, can be implemented using either gradient-free \cite{peruzzo_variational_2014,kokail_self-verifying_2019} or gradient-based methods \cite{sweke_stochastic_2019}. At first glance, the use of the former seems more effective as the evaluation of the cost function is subject to imperfection(s). However, further research developed methods of evaluating gradients analytically (i.e.~without relying on finite differences) \cite{mitarai_quantum_2018,schuld_evaluating_2019}, and the access to gradients was proven to speed up the convergence to local minima \cite{harrow_low-depth_2019}.

Similar to deep learning, quantum optimization faced a problem of vanishing gradients, or barren plateaus \cite{mcclean_barren_2018}. 
When the depth of a parametrized quantum circuit is linear in the number of qubits, it becomes a so-called $\varepsilon$-approximate $t$-design \cite{harrow_approximate_2018,brandao_local_2016}. 
For such an ensemble of unitary matrices, the expected value of $t$'th order polynomials in its entries is $\varepsilon$-close to the expected value of the same polynomials over Haar-random unitary matrices.
For variational algorithms, this implies that on average, the magnitude of gradients is exponentially small in the number of qubits. This phenomenon was first observed for VQE, although it was also reported for quantum compiling \cite{khatri_quantum-assisted_2018}, training quantum autoencoders \cite{cerezo_cost-function-dependent_2020} and quantum feedforward networks \cite{Sharma_Cerezo_Cincio_Coles_2020}. For shorter-depth quantum circuits, it was also found that the onset of barren plateaus depends on the locality of the cost function \cite{cerezo_cost-function-dependent_2020}: it was found that for cost functions consisting of local terms, the effect of barren plateaus is not as severe.

In this manuscript, we develop further on the barren plateau finding. 
Given a qubit Hamiltonian and a variational ansatz, we give a lower bound on the typical magnitude of the VQE cost function gradient, averaged over all possible assignments of the ansatz parameters.

It turns out that (i)~the variance of the gradient is a weighted sum of the variances for the individual Pauli strings comprising the Hamiltonian (i.e.\ they are independent from each other), and (ii)~the variance for an individual Pauli string can be bounded from below using the width of the \textit{causal cone} of that string. That is, when the ansatz $U$ acts on the Pauli string $h$ by conjugation, it is the number of qubits in the support of $U^\dagger h U$ which scales with the variance. Hence, the onset of barren plateaus depends not only on the locality of the Hamiltonian, but also on the structure on the ansatz.

This paper is structured as follows: in the remainder of this section, we give basic notation, introduce the idea of the barren plateaus and formulate the main results of the paper. In \S~\ref{sec:preliminaries}, we present the constructions needed for the proof: unitary designs and operators that involve averaging over said designs, which we call the mixing operators. \S~\ref{sec:proof} gives the proof of the main theorem. 
In \S~\ref{sec:numeric}, we report numerical results supporting the theoretical result. First, in \S~\ref{subsec:proximity_to_designs}, we evaluate the ability of certain families of two-qubit gates to mimic 2-designs. Then, in \S~\ref{subsec:plateau_numeric}, we directly compare the numerical distribution of derivatives with the theoretical lower bound found in this work.
\S~\ref{sec:conclusions} contains concluding remarks.

\subsection{Basic definitions and notation}

\begin{definition}[Pauli or sigma strings]
A \emph{Pauli string} 
is
a tensor product of $n$ Pauli operators $\{\mathbbm{1}, X, Y, Z \}$. The $n$-qubit identity operator $\mathbbm{1} \otimes \mathbbm{1} \otimes ... \otimes \mathbbm{1}$ is the \emph{trivial} or \textit{unit} string. The \textit{algebraic locality} or just \textit{locality} of a Pauli string is the number of non-identity Pauli matrices contained in the string.
\end{definition} 

\begin{definition}[Super Pauli strings]
If $h$ is a Pauli string, then we will call $h \otimes h$ the induced \emph{super Pauli string}. If a Pauli string acts on qubits labeled $1, 2, \dots, n$, then a super Pauli string acts on qubits labeled $1, 2, \dots, n, 1', 2', \dots, n'$.
\end{definition}

We will denote super Pauli strings as $(\sigma_1 \otimes ... \otimes \sigma_n)^{\otimes 2}$, omitting the tensor product $\otimes$ when 
there is no ambiguity. When necessary, we will mark the variables related to the second copy (on the reader's right) with an apostrophe. 

\begin{example}
A Pauli string $h = X \otimes \mathbbm{1} \otimes \mathbbm{1}$ acts nontrivially on the first out of $n = 3$ qubits. A super Pauli string $h \otimes h = (X \otimes \mathbbm{1} \otimes \mathbbm{1})^{\otimes 2}$ acts nontrivially on qubits $1$ and $1'$.
\end{example}

\begin{definition}[Ansatz]
An \textit{ansatz} $U(\boldsymbol{\theta})$ is a family of quantum circuits of fixed structure and fixed depth with gates $U_1, ..., U_p$ tunable dependent on $N$ 
parameters: $\boldsymbol{\theta} \in [0, 2 \pi)^{\times N}$. We assume that each parameter is used only in one gate, i.e.~each gate is parametrized independently. We will also assume that the gates can be grouped into what we call \textit{blocks} $G_1, ..., G_q$.
\end{definition}

We will make a few extra assumptions regarding the ansatz:

\begin{enumerate}
    \item The ansatz consists of blocks acting at most on $s$ qubits, each of which is a local 2-design (see Sec.~\ref{sec:designs});
    \item The blocks cover the ansatz in $l$ layers. Each layer acts nontrivially on all qubits, and no two blocks in the same layer act on the same qubit.
\end{enumerate}{}



\begin{definition}[Causal cone] 
Let $U$  be an ansatz, and $h$ a Pauli string. 
A block $V$ is in the \emph{causal cone} $\hat{C}(h, U)$ of $h$ under ansatz $U$, if that block cannot be eliminated from the conjugate $U^\dagger h U$. We denote as $|\hat{C}(h, U)|$ the support of this causal cone, i.e. the number of qubits on which $U^\dagger h U$ can act nontrivially.
\end{definition}{}

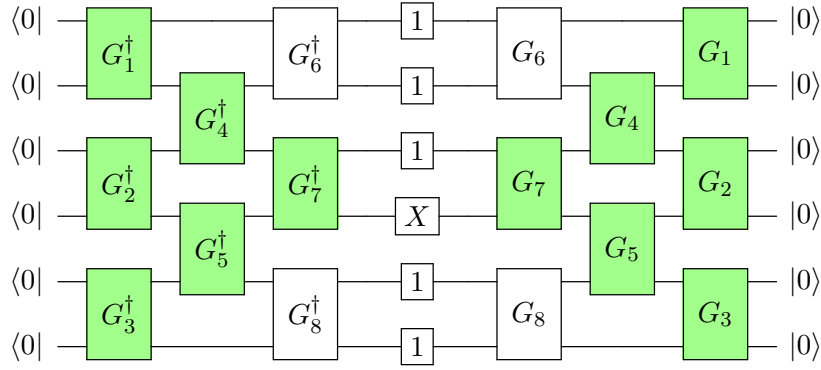
\begin{figure}
    \centering
        \begin{pgfpicture}{0em}{0em}{0.65em}{0em}
    \color{green!80!yellow!45!white}
    \pgfrect[fill]{\pgfpoint{2.3em}{-11.7em}}{\pgfpoint{2.25em}{3.12em}}
    \pgfrect[fill]{\pgfpoint{2.3em}{-7.18em}}{\pgfpoint{2.25em}{3.12em}}
    \pgfrect[fill]{\pgfpoint{2.3em}{-2.69em}}{\pgfpoint{2.25em}{3.12em}}

    \pgfrect[fill]{\pgfpoint{5.55em}{-9.45em}}{\pgfpoint{2.2em}{3.12em}}
    \pgfrect[fill]{\pgfpoint{5.55em}{-4.94em}}{\pgfpoint{2.2em}{3.12em}}
    
    \pgfrect[fill]{\pgfpoint{8.75em}{-7.18em}}{\pgfpoint{2.25em}{3.12em}}
    
    \pgfrect[fill]{\pgfpoint{16.47em}{-7.18em}}{\pgfpoint{2.25em}{3.12em}}
    
    \pgfrect[fill]{\pgfpoint{19.7em}{-9.45em}}{\pgfpoint{2.2em}{3.12em}}
    \pgfrect[fill]{\pgfpoint{19.7em}{-4.94em}}{\pgfpoint{2.2em}{3.12em}}
    
    \pgfrect[fill]{\pgfpoint{22.9em}{-11.7em}}{\pgfpoint{2.25em}{3.12em}}
     \pgfrect[fill]{\pgfpoint{22.9em}{-7.18em}}{\pgfpoint{2.25em}{3.12em}}
    \pgfrect[fill]{\pgfpoint{22.9em}{-2.69em}}{\pgfpoint{2.25em}{3.12em}}
    \end{pgfpicture}
    \mbox{
            \Qcircuit @C=1.0em @R=1.0em {
            \lstick{\bra{0}} 
            & \multigate{1}{G_1^\dagger} 
            & \qw 
            & \multigate{1}{G_6^\dagger}    
            & \qw 
            & \gate{1}
            & \qw
            & \multigate{1}{G_6} 
            & \qw 
            & \multigate{1}{G_1} 
            & \qw
            & {\ket{0}}
            \\
            \lstick{\bra{0}} 
            & \ghost{G_1^\dagger} 
            & \multigate{1}{G_4^\dagger} 
            & \ghost{G_6^\dagger} 
            & \qw  
            & \gate{1}
            & \qw 
            & \ghost{G_6} 
            & \multigate{1}{G_4} 
            & \ghost{G_1} 
            & \qw
            & {\ket{0}}
            \\
            \lstick{\bra{0}} 
            & \multigate{1}{G_2^\dagger}  
            & \ghost{G_4^\dagger}
            & \multigate{1}{G_7^\dagger} 
            & \qw  
            & \gate{1}
            & \qw 
            & \multigate{1}{G_7} 
            & \ghost{G_4}
            & \multigate{1}{G_2}  
            & \qw
            & {\ket{0}}
            \\
            \lstick{\bra{0}} 
            & \ghost{G_2^\dagger} 
            & \multigate{1}{G_5^\dagger} 
            & \ghost{G_7^\dagger}
            & \qw  
            & \gate{X}
            & \qw  
            & \ghost{G_7}
            & \multigate{1}{G_5} 
            & \ghost{G_2} 
            & \qw
            & {\ket{0}}            
            \\
            \lstick{\bra{0}} 
            & \multigate{1}{G_3^\dagger}
            & \ghost{G_5^\dagger} 
            & \multigate{1}{G_8^\dagger}
            & \qw  
            & \gate{1}
            & \qw  
            & \multigate{1}{G_8}
            & \ghost{G_5} 
            & \multigate{1}{G_3}
            & \qw
            & {\ket{0}}            
            \\
            \lstick{\bra{0}} 
            & \ghost{G_3^\dagger} 
            & \qw  
            & \ghost{G_8^\dagger}
            & \qw
            & \gate{1}
            & \qw
            & \ghost{G_8}
            & \qw  
            & \ghost{G_3} 
            & \qw
            & {\ket{0}}            
            \\
              }}
    \caption{A causal cone of a Pauli string. Highlighted gates do not cancel in $U^\dagger h U$, where $U$ is the quantum circuit pictured.}
    \label{fig:causal_cone}
\end{figure}

\begin{example}
Figure \ref{fig:causal_cone} depicts a checkerboard ansatz \cite{uvarov_machine_2019}, or alternating layered ansatz \cite{cerezo_cost-function-dependent_2020} 
acting on six qubits and consisting of three layers. Relative to a Pauli string $\mathbbm{1} \otimes \mathbbm{1} \otimes X \otimes \mathbbm{1} \otimes \mathbbm{1} \otimes \mathbbm{1}$, the causal cone for this ansatz consists of blocks $G_1$, $G_2$, $G_3$, $G_4$, $G_5$, and $G_7$. The support of this causal cone consists of all six qubits.
\end{example}{}

We will denote as $\mc{Y}$ a subset of the $n$-qubit registry, with $|\mc{Y}|$ denoting the number of qubits in that subset. Occasionally, overloading the notation, we will denote the corresponding Hilbert space with the same letter. 

{}




\subsection{Barren plateaus}

Consider a Hamiltonian $H_0$ acting on $\mc{H}$. The VQE algorithm seeks to determine the ground state of $H_0$ by preparing some parametrized ansatz state and iteratively optimizing the parameters to minimize energy. If the ansatz state is $\ket{\psi(\boldsymbol{\theta})} = U_p (\theta_p)... U_0 (\theta_0)\ket{\mathbf{0}}$, then define $E (\boldsymbol{\theta}) = \bra{\psi(\boldsymbol{\theta})} H \ket{\psi(\boldsymbol{\theta})}$ for $||\psi||_2 = 1$. Assume that each $U_a$ is a power of some skew-Hermitian operator: $U_a = e^{-\rmi \theta_a F_a}$.
We will also assume that $F_a^2 = 1$.
Suppose now that, for the optimization purposes, we take the partial derivative $\partial_a E = \frac{\partial E}{\partial \theta_a}$. Then,
\begin{equation}
    \label{eq:partial_E}
    \partial_a E = \bra{\mathbf{0}}U^\dagger_B [\rmi F_a, U^\dagger_A H_0 U_A] U_B \ket{\mathbf{0}},
\end{equation}
where $U_B = U_a \dots U_0, \ U_A = U_p \dots U_{a+1}$ for a fixed $U_a$ (with B and A standing for before and after the gate in question; $U_a$ itself can be merged into either operator).

The question posed by McClean and coworkers \cite{mcclean_barren_2018} is how $\partial_a E$ behaves on average over all possible values of $\boldsymbol{\theta}$. In answering this question, they use the theory of so-called $t$\textit{-designs}. Roughly, a set of unitaries $\mc{A} \subset U(2^n)$ is a $t$-design if $\mc{A}$ mimics certain properties of the unitary group $U(2^n)$. That is, random sampling from the unitary group $U(2^n)$ can be imitated in a certain sense by random sampling from $\mc{A}$. We will postpone the exact definition until \S~\ref{sec:designs}.

It was found that if either $U_B$ or $U_A$ forms a 1-design, 
then the average value $\langle \partial_a E \rangle_{\boldsymbol{\theta}}$ is equal to zero \cite{mcclean_barren_2018}. We will henceforth assume that the conditions of that proposition are met, and therefore $\langle \partial_a E \rangle_{\boldsymbol{\theta}} = 0$.

The barren plateau statement bounds the second moment of $\partial_a E$, i.e.~$\Var \partial_a E = \langle (\partial_a E)^2 \rangle_{\boldsymbol{\theta}}$. 
\begin{theorem}[McClean et al. \cite{mcclean_barren_2018}]
\label{thm:mcclean}
Let $U_A$ or $U_B$ form a 2-design. Then, for any Hamiltonian $H_0$ consisting of $\operatorname{poly}(n)$ Pauli strings, $\Var \partial_a E \in O(2^{-2n})$. 
\end{theorem}{}

While the fact that the expected value of the derivative is zero is not interesting in itself, the exponential upper bound on the variance means that the gradients will, on average, be very close to zero, requiring an exponential number of measurements to resolve against 
finite sampling effects and noise.

Since it is known that a random parallel circuit can form an approximate 2-design using $O(n)$ layers \cite{harrow_approximate_2018,brandao_local_2016}, this result suggests that VQE will, on average, take an exponential number of measurements to converge if the underlying circuit is linear in depth. 
However, for logarithmic depth circuits, this is not necessarily the case. 
In particular, as was found by Cerezo et al. \cite{cerezo_cost-function-dependent_2020}, the result depends on the structure of the cost function, $H_0$. 

\subsection{Statement of main results}

\begin{theorem}
\label{thm:main}
Let $H$ be an $n$-qubit Hamiltonian consisting of Pauli strings $h_i$: $H = \sum c_i h_i$ with finite $c_i \in \mathbb{R}$. Let the ansatz $U$ consist of $l$ layers, and denote $l_c$ the layer which contains the block $G_k$ depending on parameter $\theta_a$. 
Let each block of the ansatz be an independently parametrized local 2-design. Let the block $G$ also be decomposable into $G = G_A e^{-i \theta_a F} G_B$, where $G_A$ and $G_B$ are local 2-designs not depending on $\theta_a$. Then, the variance of the gradient of $E$ with respect to that parameter is bounded below as follows:

\begin{equation}
    \Var \partial_a E \geq \frac{2 \cdot 4^{|\mc{Y}_k|}}{4^{|\mc{Y}_k|} - 1} \left( \frac34 \right)^{l - l_c}  \sum_i c_i^2 \cdot 3^{-|\hat{C}(h_i, U)|},
\end{equation}{}
where $|\hat{C}(h_j, U)|$ is the number of qubits in the causal cone of the $j^{th}$ Pauli string, and the summation is over those Pauli strings whose causal cone contains the block $G$.
\end{theorem}{}


The proof of this theorem uses the fact that $\partial_a E (h_i)$ are uncorrelated random variables:

\begin{lemma}
\label{lemma:decouple}
In the conditions of Theorem \ref{thm:main}, the individual Pauli string coefficients make independent contributions to the total variance:
\begin{equation}
        \Var \partial_a E (H) = \sum_i c_i^2 \Var \partial_a E (h_i).
\end{equation}
\end{lemma}




Our results imply that algebraic locality is not the only factor in the emergence of barren plateaus for a given Pauli string $h$. More important is the maximum possible locality under conjugation with the ansatz: $h \mapsto U^\dagger h U$. Informally, if the causal cone of a Pauli string has many qubits in its support, this string may be difficult to optimize by gradient descent. 




\section{Preliminaries}
\label{sec:preliminaries}

\subsection{Idea of the proof}

To estimate the variance, we will need to integrate $(\partial_aE) ^2$ over all possible assignments of the parameters $\boldsymbol{\theta}$. We consider the expression $(\partial_a E)^2$ in the Heisenberg picture, that is, we think of all operators as acting on $H \otimes H$. For example, the first such operator maps $H \otimes H$ to $(U_p^\dagger \otimes U_p^\dagger) (H \otimes H) (U_p \otimes U_p)$. Since we assumed that the blocks are parametrized independently, we can also take their expected values independently. If $U_p$ depends on parameters $\boldsymbol{\theta}_j =  \theta_{j_1}, ..., \theta_{j_m}$ for some $j_1, ..., j_m$, then the operator we care about takes the following form:

\begin{equation}
\label{eq:true_mixer}
M = \int (U_p^\dagger \otimes U_p^\dagger) (\star) (U_p \otimes U_p) 
\mathrm{d} \boldsymbol{\theta}_j
= \int 
\adjustbox{raise=31pt}{
\Qcircuit @C=1em @R=.7em 
{& \multigate{1}{U^\dagger} &  \multigate{3}{\star} 
& \multigate{1}{U} & \qw
\\
& \ghost{U^\dagger} & \ghost{\star}
& \ghost{U} & \qw
\\
& \multigate{1}{U^\dagger} & \ghost{\star}
& \multigate{1}{U} & \qw
\\
& \ghost{U^\dagger} & \ghost{\star}
& \ghost{U} & \qw
}
}
\mathrm{d} \boldsymbol{\theta}_j
\end{equation}
where the star ($\star$) is a placeholder for a Hermitian operator on $\mc{H} \otimes \mc{H}$. The commutator found in \eqref{eq:partial_E} can also be viewed as a superoperator $[\rmi F, \star]^{\otimes 2}$. Graphically, we can express this superoperator like this:

\newcommand{\legw}{1}
\newcommand{\gapw}{2}
\newcommand{\gaph}{1}
\newcommand{\barh}{0.6}
\newcommand{\wireh}{0.2}
\newcommand{\wiregap}{0.5}
\newcommand{\wirel}{0.3}

\begin{equation}
    [\rmi F, \star] =
    \adjustbox{raise=-7pt}{
    \begin{tikzpicture}[thick,scale=0.5]
    \node[blank] at (\gapw/2 + \legw, \gaph/2) {$\star$};

    \draw 
    (0, 0) 
    -- (0, \gaph + \barh) 
    -- (\gapw + \legw + \legw,\gaph + \barh)
    -- (\gapw + \legw + \legw,0)
    -- (\gapw + \legw,0)
    -- (\gapw + \legw, \gaph)
    -- (\legw, \gaph)
    -- (\legw,0)
    -- (0, 0)
    
    (0, \wireh) -- (-\wirel, \wireh)
    (0, \wireh + \wiregap) -- (-\wirel, \wireh + \wiregap)
    
    (\legw, \wireh) -- (\legw + \wirel, \wireh)
    (\legw, \wireh + \wiregap) -- (\legw + \wirel, \wireh + \wiregap)
    
    (\legw + \gapw, \wireh) -- (\legw + \gapw - \wirel, \wireh)
    (\legw + \gapw, \wireh + \wiregap) -- (\legw + \gapw - \wirel, \wireh + \wiregap)
    
    (\gapw + \legw + \legw, \wireh) -- (\gapw + \legw + \legw + \wirel, \wireh)
    (\gapw + \legw + \legw, \wireh + \wiregap) -- (\gapw + \legw + \legw + \wirel, \wireh + \wiregap)
    
    ;
    \end{tikzpicture}
    }
\end{equation}

\begin{figure}
    \begin{tikzpicture}
    \node[] (formula) {$\Var \partial_a E = \displaystyle{\int} \mathrm{d} \boldsymbol{\theta}$};
    \node (fig1) [above right of =formula, xshift=6.5cm, yshift=0.8cm] {\includegraphics[width=0.7\linewidth]{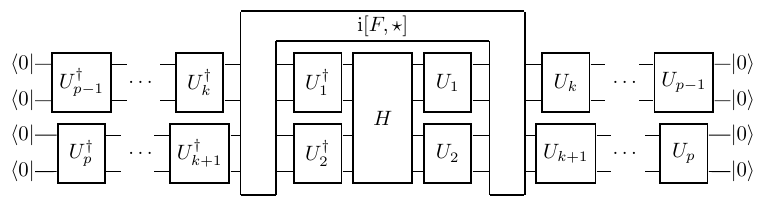}};
    \node (fig2) [below right of =formula, xshift=6.5cm, yshift=-0.8cm] {\includegraphics[width=0.7\linewidth]{vardE.pdf}};
    \end{tikzpicture}

    \caption{Variance of the derivative of $E$, expressed as an integration over all possible assignments of $\boldsymbol{\theta}$. 
    }
    \label{fig:variance_as_diagram}
\end{figure}

In this graphical language, the value of $\operatorname{Var} \partial_a E$ is expressed as a diagram shown in Fig.~\ref{fig:variance_as_diagram}.

Instead of evaluating the action of operators like $M$ for a specific ansatz, we instead assume that each block of the ansatz constitutes a local 2-design, in which case one can compute their action exactly. We will refer to such operators as ``mixing operators''.


The action $[\rmi F, \star]^{\otimes 2}$ can be written down explicitly using the assumption that there are two mixing operators around it. In which case, its role reduces to eliminating those Pauli strings that don't share support with $F$, and multiplying all other strings by a constant. We estimate the number of strings that survive this operation by tracing a path along the structure of the ansatz.

After all these operators, we end up with a number of Pauli strings with some coefficients. Taking the expectation w.r.t.\ the zero kets eliminates those strings that contain $X$ or $Y$ Pauli matrices. We estimate the share of Pauli strings that are not eliminated in the process. The sum of their coefficients is the final value that we are after.

In the following sections, we will first derive the necessary properties of the mixing operators and of the commutator-induced superoperator. Then we will use them to prove the main statement.

\subsection{Unitary designs}
\label{sec:designs}

Let $\mc{H} = (\mathbb{C}^2)^{\otimes n}$ be the Hilbert space of $n$ qubits. The Haar measure \cite{watrous_theory_2018} $d \mu$ on the unitary group  $U(2^n)$ is the left-invariant probability measure. For any Borel set $\mc{A} \subset U(2^n)$, its measure is preserved under any unitary shift $V$: $\mu(\mc{A}) =\mu(V \mc{A})$. This measure is a natural generalization of the idea of a uniform distribution. To see this, observe that for $U(1) \equiv \{e^{\rmi \varphi} | \varphi \in [0, 2 \pi) \}$ the Haar measure is just the uniform measure on the circle of unit length. Finally, in the limit of large circuit depth, random circuits converge to the Haar measure, although the convergence in exponentially slow in the number of qubits \cite{emerson_convergence_2005}.

The Haar measure can be approximated by unitary ensembles known as unitary $t$-designs. A probability distribution $\nu$ on the unitary group $U(2^n)$ is a  \textit{unitary} $t$\textit{-design} 
if the expected value of any polynomial of power $t$ in the entries of $U$ and $U^*$ with respect to $\nu$ is the same as that w.r.t.~the Haar measure on $U(2^n)$.

Let $(\star)$ be a placeholder for a linear operator on $\mc{H}^{\otimes t}$. Then the integral of $ (U^\dagger)^{\otimes t}  \ (\star) \ U^{\otimes t} $ over the Haar-distributed $U$ is equal to a linear combination of operators permuting the $t$ copies of $\mc{H}$, with the coefficients being given by the Weingarten function of the permutations\cite{collins_integration_2006,Samuel_1980}. For $t = 2$, these permutations are the identity $\mathbbm{1}$ and the swap $\mc{S}$. 


\begin{lemma}
\label{lemma:int_haar}
Let $d \mu(U)$ be the Haar measure on $U(2^n)$.
For every Hermitian $A, B$ acting on $n$ qubits, we have the following \cite{Samuel_1980}:

\begin{equation}
\label{eq:int_haar_first}
    \int  (U^\dagger) A U d\mu(U) = \frac{\Tr A}{\dim U} \mathbbm{1}.
\end{equation}{}

\begin{equation}
\label{eq:int_haar}
\begin{aligned}
    &\int (U^\dagger)^{\otimes 2} (A \otimes B) U^{\otimes 2} d\mu(U) = \\
    & = \frac{1}{(\dim U)^2 - 1} \left[  
    (\Tr A \Tr B  - \frac{1}{\dim U} \Tr AB) \mathbbm{1} \otimes  \mathbbm{1} + \right. \\
    & \left. + (\Tr AB  - \frac{1}{\dim U} \Tr A \Tr B) \mc{S} \right].
\end{aligned}{}   
\end{equation}{}
\end{lemma}{}

We can also rewrite the identity (\ref{eq:int_haar}) for the inputs that are not tensor factorizable. Observe that $\Tr (AB) = \Tr \mc{S} (A \otimes B)$. Now, $\Tr A \Tr B = \Tr (A \otimes B)$, 
so the expression \eqref{eq:int_haar} is a function of $(A \otimes B)$ and not of $A$ or $B$ independently.
Using the linearity of trace and swap, we can extend this identity:

\begin{corollary}
Let $C \in \mathrm{Herm}(\mathcal{H} \otimes \mc{H})$. Then, under the conditions of Lemma \ref{lemma:int_haar}, we have established the following:

\begin{equation}
\label{eq:int_haar_swap}
\begin{aligned}
    &\int (U^\dagger)^{\otimes 2} C U^{\otimes 2} d\mu(U) = \\
    & = \frac{1}{(\dim U)^2 - 1} \left[  
    (\Tr C  - \frac{1}{\dim U} \Tr \mc{S} C) \mathbbm{1} \otimes  \mathbbm{1} + \right. \\
    & \left. + (\Tr \mc{S} C  - \frac{1}{\dim U} \Tr C) \mc{S} \right].
\end{aligned}{}   
\end{equation}{}
\end{corollary}{}

\subsection{Local mixing operator}
\label{subsec:local_mixer}

When we assumed that each block of the ansatz is an independent 2-design, we implied a certain structure on the ansatz $U$. That is, an integral of any function $f(U)$ over the instances of $U$ decomposes into a multiple integral over the instances of blocks $G_k$:

\begin{equation}
    \int f(U) d\mu (U) = \int \dots \int f(U) d\mu(G_1) ... d\mu(G_q).
\end{equation}{}

Hence, a superoperator $\int (U^\dagger)^{\otimes t} \star U^{\otimes t}  d \mu(U)$ present in formulas (\ref{eq:int_haar_first}), (\ref{eq:int_haar}) is also replaced by a multitude of local superoperators which we define below for $t=2$.

\begin{definition}

Let $\mc{Y}$ be a subset of the qubit registry. Define the \textit{local mixing operator}, or simply a \textit{mixer}  $M_{\mc{Y}}: \mc{L} (\mc{H}\otimes \mc{H}) \rightarrow \mc{L} (\mc{H}\otimes \mc{H})$ as follows:
\begin{equation}
\begin{aligned}
    \label{eq:m2}
    & M_{ \mc{Y}} (h_1 \otimes h_2) = \int d\mu_{\mc{Y}} (U)
    (U^\dagger \otimes U^\dagger)
    (h_1 \otimes h_2)
    (U \otimes U),
\end{aligned}{}
\end{equation}{}
where $\mu_{\mc{Y}}$ is a Haar distribution of unitaries acting nontrivially on $\mc{Y}$.

\end{definition}{}

\begin{proposition}
\label{prop:m2_decomposed}
Let $h$ be a Pauli string. If its substring $h_{\mc{Y}}$ is nontrivial, then

\begin{equation}
    M_{\mc{Y}}(h \otimes h) = \frac{1}{4^{|\mc{Y}|} - 1} \left( \sum_{\sigma_\mc{Y} \neq \mathbbm{1}} (\sigma_{\mc{Y}} \otimes h_{\mc{H} \setminus \mc{Y}})^{\otimes 2}\right),
\end{equation}{}
where the summation extends over all nontrivial Pauli substrings $\sigma_{\mc{Y}}$.
Otherwise,  $M_{\mc{Y}}(h \otimes h) = h \otimes h$.

\end{proposition}{}
\begin{proof}
We first apply formula (\ref{eq:int_haar}):
\begin{equation}
\begin{aligned}
\label{eq:m2_proof_1}
    & M_{\mc{Y}}(h \otimes h) = \\
    & = \frac{1}{4^{|\mc{Y}|} - 1} (h \otimes h)_{\mc{H} \setminus \mc{Y}} \otimes \left[  
    \left(\Tr (h \otimes h)_{\mc{Y}}  - \frac{1}{2^{|\mc{Y}|}} \Tr \mc{S}_{\mc{Y}} (h \otimes h)_{\mc{Y}} \right) \mathbbm{1} \otimes  \mathbbm{1} + \right. \\
    & \left. + \left(\Tr \mc{S}_{\mc{Y}} (h \otimes h)_{\mc{Y}}  - \frac{1}{2^{|\mc{Y}|}} \Tr (h \otimes h)_{\mc{Y}}\right) \mc{S}_\mc{Y} \right],
\end{aligned}{} 
\end{equation}
where $\mc{S}_\mc{Y}$ is the swap operator permuting pairs of qubits $(i, i')$ for $i \in \mc{Y}$. It means that $\mc{S}_\mc{Y}$ is a tensor product of two-qubit swap gates $\mc{S}_2$. 
First, note that $\Tr \mc{S}_{\mc{Y}} (h \otimes h)_{\mc{Y}} = 2^{|\mc{Y}|}$ and $\Tr (h \otimes h)_{\mc{Y}}$ is equal to zero for nontrivial $h_\mc{Y}$ and $4^{|\mc{Y}|}$ for a trivial substring. Next, $\mathcal{S}_2$ is decomposed as 
\begin{equation}
\label{eq:swap_decomp}
    \mathcal{S}_2 = \frac{1}{2}\left(X \otimes X + Y \otimes Y + Z \otimes Z + \mathbbm{1} \otimes \mathbbm{1} \right).
\end{equation}
Applying this decomposition to $\mc{S}_\mc{Y}$ yields a sum of all possible super Pauli strings (including the trivial string):
\begin{equation}
    \mc{S}_{\mc{Y}} = \frac{1}{2^\mc{|Y|}} \sum_{\sigma_1, ..., \sigma_\mc{|Y|}}
    (\sigma_1 \otimes \sigma_2 \otimes ... \otimes \sigma_\mc{|Y|})^{\otimes 2}.
\end{equation}
Substituting this decomposition into \eqref{eq:m2_proof_1} we recover the desired result.
\end{proof}{}

The following proposition will later help establish that distinct Pauli strings decouple:

\begin{proposition}
\label{prop:paulis_decouple}
Let $\mc{Y}_1, ..., \mc{Y}_N$ be a collection of qubit subsets,
such that $\mc{Y}_1 \cup ... \cup \mc{Y}_N$ contains all $n$ qubits (the subsets are allowed to intersect). Let $h_1, h_2$ be two distinct Pauli strings. Then, 
$M_{ \mc{Y}_N} \circ \dots \circ  M_{ \mc{Y}_1} (h_1 \otimes h_2) = 0.$
\end{proposition}{}
\begin{proof}
Since $\mc{Y}_1 \cup ... \cup \mc{Y}_N$ is required to contain all qubits, there is necessarily a subset of qubits $\mc{Y}_j$ on which $h_1$ and $h_2$ act differently. 
More specifically, let the qubit $q$ be one where $h_1$ and $h_2$ differ. 
Let us first assume that $\mc{Y}_j $ does not overlap with any preceding subsets. After the action of $M_{\mc{Y}_j}$ (see (\ref{eq:int_haar}),(\ref{eq:m2})), all terms are proportional to either $\Tr_\mc{Y, Y'} (h_1 \otimes h_2)$ or $\Tr_\mc{Y, Y'} \mc{S_\mc{Y, Y'}} (h_1 \otimes h_2)$. The Pauli matrices are traceless, so $\Tr_\mc{Y, Y'} (h_1 \otimes h_2) = 0$. However, if a swap acts on qubit $q$, then $\Tr_{q, q'} \mc{S}_{q, q'} (h_1 \otimes h_2)|_{q, q'} = \Tr (\sigma_{1_q} \sigma_{2_q})$, which is nonzero if and only if the Pauli matrices $\sigma_{1_q}, \sigma_{2_q}$ are equal.

Finally, if $\mc{Y}_j$ does overlap with some preceding subset, then by Proposition \ref{prop:m2_decomposed} the input of $M_{\mc{Y}_j}$ consists of a sum of Pauli string pairs, all still distinct in qubits $q, q'$. All of the reasoning above still applies, and the result is zero.
\end{proof}{}

\subsection{Commutator}

Let $F$ be a Hermitian operator that acts on a subset of qubits $\mc{Y}$, then $\rmi [F, \star]$ is a superoperator on $\mathrm{Herm}(\mc{H})$.
We also introduce 
$\mc{C}_{\mc{Y}} = (\rmi[F, \star] \otimes \rmi[F, \star]): \mc{L}(\mc{H} \otimes \mc{H}) \rightarrow \mc{L}(\mc{H} \otimes \mc{H})$, where each commutator acts on a copy of $\mc{L}(\mc{H})$.

Recall our assumption that the before and after parts of the block $G_B, G_A$ are independent 2-designs, i.e. that the gate depending on $\theta$ sits somewhere in the middle of the block. 
This means that we will be interested in the behavior of a commutator operator sandwiched between two local mixing operators with the same support.

\begin{proposition}
\label{prop:commutator}
The following identities hold:
\begin{enumerate}
    \item For every $F \in \mathrm{Herm}(\mc{Y})$, $\mc{C}_{\mc{Y}}(\mathbbm{1}^{\otimes |\mc{Y}|} \otimes \mathbbm{1}^{ \otimes |\mc{Y}|})$ vanishes. Thus, $M_{\mc{Y}} \circ \mc{C}_{\mc{Y}} \circ M_{\mc{Y}} (\mathbbm{1}^{\otimes |\mc{Y}|} \otimes \mathbbm{1}^{\otimes |\mc{Y}|})=0$.
    \item Let $F$ be a nontrivial Pauli string acting on $\mc{Y}$. Then, for any nontrivial Pauli string $h$ acting on $\mc{Y}$
    \begin{equation}
        \label{eq:sandwiched_commutator}
         M_{\mc{Y}} \circ \mc{C}_{\mc{Y}} \circ M_{\mc{Y}} \left( h^{\otimes 2}\right) = \frac{2 \cdot 4^{|\mc{Y}|}}{4^{|\mc{Y}|} - 1} M\left( h^{\otimes 2}\right).
    \end{equation}
\end{enumerate}{}
\end{proposition}{}

\begin{proof}
The first part follows directly: identity operator commutes with any other operator.
To prove the second part, we will sequentially apply the operators in the left-hand side of \eqref{eq:sandwiched_commutator}. First, the local mixing operator returns a linear combination of all nontrivial Pauli strings $\boldsymbol{\sigma}_i$: $M_{\mc{Y}} \left( h^{\otimes 2}\right) = 1/ (4^{|\mc{Y}|} - 1)\sum \boldsymbol{\sigma}_i^{\otimes 2}$. After applying $\mc{C}_{\mc{Y}}$ to each super Pauli string we either get zero for those commuting with $F$ and some other super Pauli string $\boldsymbol{\kappa}_i \otimes \boldsymbol{\kappa}_i$ multiplied by 4 for those anticommuting with $F$:
\begin{equation}
    ([\rmi F, \boldsymbol{\sigma}_i])^{\otimes 2} = (\pm 2 \boldsymbol{\kappa}_i)^{\otimes 2} = 4 \boldsymbol{\kappa}_i^{\otimes 2}.
\end{equation}

For any nontrivial Pauli string $F$, there are exactly $4^{|\mc{Y}|} / 2$ nontrivial Pauli strings that anticommute with $F$. 
Indeed, let $F$ contain $m$ nontrivial Pauli matrices and let $P$ be some Pauli string that we wish to construct, so that it anticommutes with $F$. How many ways of constructing $P$ are there? There must be an odd number of sites $j$ such that Pauli matrices $F_j$ and $P_j$ are both nontrivial and not equal to each other. We can pick such sites in $2^{m - 1}$ ways. Then, for each of these sites, there is a choice of 2 Pauli matrices not commuting with $F_j$. For all sites where $F_j$ is nontrivial, but which are not included in our selection, $P_j$ is either equal to $\mathbbm{1}$ or to $F_j$. Finally, in all sites where $F_k = \mathbbm{1}$, we are free to choose any Pauli matrix. Hence, when the choice of sites is fixed, we have $2^m 4^{|\mc{Y}| - m}$ options. Multiplying this by $2^{m - 1}$, we get $4^{|\mc{Y}|} / 2$.

Overall, the result is the following: the first mixer produces a sum of all possible nontrivial super Pauli strings, the commutator $\mc{C}$ keeps $4^{|\mc{Y}|} / 2$ of them and multiplies them by 4, and then the second mixer again turns each super string into a sum of all possible super Pauli strings. Collecting the prefactors yields \eqref{eq:sandwiched_commutator}.

\end{proof}

\begin{remark}
For some ans\"atze used in the numerical experiments (Section \ref{sec:numeric}), the requirement that the gate $\exp{\mathrm{i}\theta F}$ should be surrounded by independent local 2-designs is not fulfilled. However, as long as preceding and successive blocks constitute approximate 2-designs, the asymptotic behavior is not distorted: the key observation is that the commutator receives a collection of all possible Pauli strings, with uniform weights. The blocks that are not surrounded by other blocks (i.e. those in the first and in the last layer) do not appear to alter the asymptotic behavior (see Section \ref{sec:numeric}).
\end{remark}


\section{Proof of Theorem \ref{thm:main}}
\label{sec:proof}

Following the definitions, the variance can be expressed as an average of a certain operator over the zero ket vectors 
$\ket{\boldsymbol{00}} = (\ket{\boldsymbol{0}} \otimes \ket{\boldsymbol{0}})$. To write down that operator, we use the assumption that individual blocks $G_1, ..., G_q$ are local 2-designs, and replace the integration with local mixing operators $M_{\mc{Y}_1} ... M_{\mc{Y}_q}$:
\begin{equation}
\label{eq:all_mixers}
\Var \partial_a E (H) 
= \bra{\boldsymbol{00}}
M_{ \mc{Y}_1} \circ 
\dots \circ  M_{ \mc{Y}_k} \circ \mc{C} \circ M_{ \mc{Y}_k} \circ 
\dots \circ  M_{ \mc{Y}_q} (H \otimes H) \ket{\boldsymbol{00}}.
\end{equation}{}
Note the reverse order of the mixing operators: if the ansatz state is $\ket{\psi} = G_q ... G_1 \ket{\mathbf{0}}$, then $\bra{\psi} H \ket{\psi} = 
\bra{\mathbf{0}} G_1^\dagger ... G_q^\dagger H G_q ... G_1 \ket{\mathbf{0}}$. 
Therefore, in the Heisenberg picture, the conjugation with unitaries reverses.

The action of all operators in the right hand side of \eqref{eq:all_mixers}
is linear, so we can replace $H \otimes H$ by $\sum_{i, j}c_i c_j h_i \otimes h_j$. From Proposition \ref{prop:paulis_decouple} we know that all terms with $i \neq j$ will vanish, so we arrive at
\begin{equation}
    \label{eq:paulis_decouple}
    \Var \partial_a E (H) = \sum_i c_i^2 \Var \partial_a E (h_i),
\end{equation}
effectively reducing the problem to the case when the Hamiltonian of interest is a single Pauli string $h$.

In the next three subsections, we estimate the value of $\Var \partial_a E (H)$ by sequentially applying the superoperators shown in \eqref{eq:all_mixers}.

\subsection{First portion of mixing operators}

Let us now follow the evolution of some Pauli string $h_i$ along the application of the mixing operators before the commutation operator $\mc{C}$. Each mixing operator replaces the super Pauli string with the sum of super Pauli strings with all possible nontrivial substrings in its support. For example, a two qubit mixing operator takes one super Pauli string and returns 15 super Pauli strings (see Proposition \ref{prop:m2_decomposed}):

\begin{equation}
    M\left((X \otimes X)^{\otimes 2} \right) = \frac{1}{15}
    \left( \sum_{i,j} (\sigma_i \otimes \sigma_j)^{\otimes 2}
    + \sum_{i} \left((\sigma_i \otimes \mathbbm{1})^{\otimes 2}
    + (\mathbbm{1} \otimes \sigma_i)^{\otimes 2} \right)
    \right).
\end{equation}

After a sequence of such mixing operators, the super Pauli string $h \otimes h$ is transformed into a sum of some other super Pauli strings $g_\alpha$: $M_{\mc{Y}_k} \circ ... \circ M_{\mc{Y}_q} (h_i \otimes h_i) = \sum c'_\alpha g_\alpha \otimes g_\alpha$. This collection is rather difficult to describe, however, from the properties of the mixers we know that (i) the coefficients $c'_\alpha$ sum up to one, and that (ii) the support of every Pauli string $g_\alpha$ is bounded by the support of the causal cone $|\hat{C}(h, U)|$. Also, one can show that every qubit in the causal cone is in the support of some Pauli string $g_\alpha$.




\subsection{Elimination of terms by commutator}

The block $G_k$ corresponds to a triple of operators $M_{ \mc{Y}_k} \circ \mc{C} \circ M_{ \mc{Y}_k}$, of which we already used the first one. The pair of operators $M_{\mc{Y}_k} \mc{C}$ now acts in the following way: the strings whose domain does not intersect that of $\mc{C}$ are eliminated, while all other strings are multiplied by a constant coefficient depending on the size of the block containing $\mc{C}$ (see Proposition \ref{prop:commutator}). To bound the number of surviving terms from below, we explicitly track a subset of such terms. 

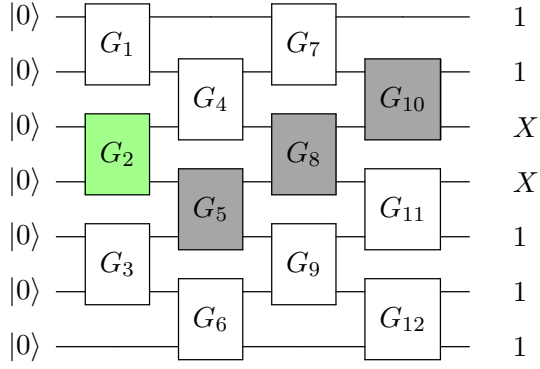
\begin{figure}
    \centering
    \begin{pgfpicture}{0em}{0em}{0em}{0em}
    \color{gray!70}
    \pgfrect[fill]{\pgfpoint{11.32em}{-4.25em}}{\pgfpoint{2.65em}{2.8em}}
    \pgfrect[fill]{\pgfpoint{8.1em}{-6.15em}}{\pgfpoint{2.25em}{2.8em}}
    \pgfrect[fill]{\pgfpoint{4.9em}{-8.05em}}{\pgfpoint{2.2em}{2.8em}}
    \color{green!80!yellow!45!white}
    \pgfrect[fill]{\pgfpoint{1.65em}{-6.15em}}{\pgfpoint{2.25em}{2.8em}}
    \end{pgfpicture}
    \mbox{
            \Qcircuit @C=1.0em @R=1.0em {
            \lstick{\ket{0}} 
            & \multigate{1}{G_1} 
            & \qw 
            & \multigate{1}{G_7}    
            & \qw 
            & \qw
            & \rstick{1}
            \\
            \lstick{\ket{0}} 
            & \ghost{G_1} 
            & \multigate{1}{G_4} 
            & \ghost{G_7} 
            & \multigate{1}{G_{10}} 
            & \qw  
            & \rstick{1}
            \\
            \lstick{\ket{0}} 
            & \multigate{1}{G_2}  
            & \ghost{G_4}
            & \multigate{1}{G_8} 
            & \ghost{G_{10}}
            & \qw  
            & \rstick{X}
            \\
            \lstick{\ket{0}} 
            & \ghost{G_2} 
            & \multigate{1}{G_5} 
            & \ghost{G_8}
            & \multigate{1}{G_{11}} 
            & \qw  
            & \rstick{X}
            \\
            \lstick{\ket{0}} 
            & \multigate{1}{G_3}
            & \ghost{G_5} 
            & \multigate{1}{G_9}
            & \ghost{G_{11}} 
            & \qw  
            & \rstick{1}
            \\
            \lstick{\ket{0}} 
            & \ghost{G_3} 
            & \multigate{1}{G_6}
            & \ghost{G_8}
            & \multigate{1}{G_{12}}
            & \qw
            & \rstick{1}
            \\
            \lstick{\ket{0}} 
            & \qw
            & \ghost{G_6}
            & \qw
            & \ghost{G_{12}}
            & \qw
            & \rstick{1}
            \\
              }
        }
    \caption{An example of a path constructed out of blocks with nontrivial inputs. Dark blocks correspond to mixing operators, highlighted block contains the commutator.}
    \label{fig:comm_path}
\end{figure}

If the block $G_k$ is outside the causal cone of $h \otimes h$, then the derivative $\partial_\theta E$ is trivially zero. 
So, excluding this case, let us assume that the block $G_k$ is within the causal cone of $h$. This means that we can find a sequence of blocks $G_{j_l}, ... G_{j_{l_c + 1}}$ situated in layers $l, ..., l_c + 1$, such that the first one shares support with $h$, and each $G_{j_{k-1}}$ shares support with $G_{j_k}$. Finally, we require that $G_{j_{l_c + 1}}$ shares support with the block $G_k$. Thus, we have established a causal path from $h$ to the commutator (see Fig. \ref{fig:comm_path}). 

Now, the output of the mixer $M_{j_l}$ contains $4^{|\mc{Y}_{j_l}|} - 1$ super Pauli strings with equal coefficients, at least $3/4$ of which act nontrivially in the support of $G_{j_{l-1}}$. For example, if $G_{j_l}$ is a two-qubit block, it outputs 15 super Pauli strings, of which 12 share domain with the next block $G_{j_{l - 1}}$. The next mixer 
$M_{j_{l - 1}}$ takes those super Pauli strings, and for each of them, outputs $4^{|\mc{Y}_{j_{l - 1}}|} - 1$ super Pauli strings, of which at least $3/4$ again have nontrivial action in the support of the next block. Continuing on, we find that the total weight of such super Pauli strings is at least $(3/4)^{l - l_c}$. Then it gets multiplied by $\frac{2 \cdot 4^{|\mc{Y}_k|}}{4^{|\mc{Y}_k|} - 1}$. 

Note that the blocks outside the specified path can increase this number, but not decrease it. For example, in the notation of Fig. \ref{fig:comm_path}, we first acted with the mixer $M_{10}$, corresponding to the block $G_{10}$, and got a collection of super Pauli strings in the output, some of which act nontrivially on qubits 3 and 4, the support of $G_8$. The action of the mixer $M_{11}$ cannot make those strings lose the nontrivial action on that support. In principle, such block could bring more strings to act nontrivially there, but for a lower bound this is not important.

{}

\subsection{Second portion of the mixing operators}

Let us apply all the remaining operators except the set of mixers $B = \{ M_{\mc{Y}_{1}}, ..., M_{\mc{Y}_m} \}$ corresponding to the first layer of the ansatz. Upon doing that, we get a linear combination of super Pauli strings $\sum_\alpha c''_\alpha g'_\alpha \otimes  g'_\alpha$. The coefficients $c''_\alpha$ sum up to at least $\frac{2 \cdot 4^{|\mc{Y}_k|}}{4^{|\mc{Y}_k|} - 1} \cdot (3/4)^{l - l_c} $. Each super string $g'_\alpha \otimes  g'_\alpha$ will then go through this layer of mixers and then the output will get averaged over the zero ket vector $\ket{\boldsymbol{00}}$.

For every $g'_\alpha \otimes  g'_\alpha$, the number resulting from this series of operations is greater or equal to $\prod_{M_{\mc{Y}} \in B} \frac{2^{|\mc{Y}|} - 1}{4^{|\mc{Y}|} - 1}$. The super Pauli string $g'_\alpha \otimes  g'_\alpha$ can act trivially or nontrivially on the support of each $M_{\mc{Y}} \in B$. If it acts trivially, then $M_{\mc{Y}}$ does nothing. In the opposite case, $M_{\mc{Y}}$ yields $4^{|\mc{Y}|} - 1$ super Pauli strings with equal weights. Of these strings, only $2^{|\mc{Y}|} - 1$ consist entirely of identity matrices and Pauli $Z$ matrices, and only these strings yield a 1 when averaged over $\ket{\boldsymbol{00}}$. Repeating for all mixers in $B$ yields the lower bound $\prod_{M_{\mc{Y}} \in B} \frac{2^{|\mc{Y}|} - 1}{4^{|\mc{Y}|} - 1}$.


Summing up all of the above, we write the lower bound on $\Var \partial_\theta E(h_i)$:
\begin{equation}
\label{eq:var_theta}
    \partial_\theta E (h) \geq \frac{2 \cdot 4^{|\mc{Y}_k|}}{4^{|\mc{Y}_k|} - 1} \left( \frac34 \right)^{l - l_c} 
    \prod_{M_{\mc{Y}} \in B} 
    \frac{2^{|\mc{Y}|} - 1}{4^{|\mc{Y}|} - 1} = 
    \frac{2 \cdot 4^{|\mc{Y}_k|}}{4^{|\mc{Y}_k|} - 1} \left( \frac34 \right)^{l - l_c}
    \prod_{M_{\mc{Y}} \in B} 
    \frac{1}{2^{|\mc{Y}|} + 1}.
\end{equation}{}

The last product can be bounded from below by $3^{-|\hat{C}(h, U)|}$. Here is how: $1/ (2^{|\mc{Y}|} + 1) = (1/2^{|\mc{Y}|}) \cdot 1/(1 + 2^{-|\mc{Y}|})$. Since $|\mc{Y}| \geq 1$, then the second factor of the right-hand side of this equality is greater or equal than $2/3$. The product can be then transformed as follows:

\begin{equation}
\label{eq:massage_var_theta}
     \prod_{M_{\mc{Y}} \in B} 
    \frac{1}{2^{|\mc{Y}|} + 1} \geq      
    \prod_{M_{\mc{Y}} \in B}  \frac{1}{2^{|\mc{Y}|}} \frac{2}{3}
    = \left( \frac{2}{3} \right)^{|B|} \frac{1}{2^{|\hat{C}(h, U)|}}.
\end{equation}{}
Now observe that the number of blocks is not greater than the number of qubits, and hence the last part of \eqref{eq:massage_var_theta} is lower bounded by $3^{-|\hat{C}(h, U)|}$, which concludes the proof.







\section{Numerical experiments}
\label{sec:numeric}

\subsection{Proximity of local blocks to 2-designs}
\label{subsec:proximity_to_designs}


It is known that approximate 2-designs can be prepared by a polynomial depth random circuit \cite{harrow_approximate_2018,brandao_local_2016}. However, here we are interested in local blocks whose properties are not guaranteed by asymptotic estimates.

We performed a series of numerical experiments to compare certain two-qubit blocks to exact unitary designs. A simple way of evaluating the proximity of the gate families to the Haar measure is to measure the distance to the so-called quantum $t$-tensor product expander (TPE) \cite{brandao_local_2016,Low_2010}. A family of random unitary gates $\nu$ is a $\lambda$-approximate TPE if $||\mathbb{E}_{Haar} (U^{\otimes t} \otimes (U^*)^{\otimes t}) - \mathbb{E}_\nu (U^{\otimes t} \otimes (U^*)^{\otimes t}) ||_p \leq \lambda$ for $p=\infty$. 
The trace definition of an approximate $t$-design involves the same quantity for $p = 1$. 
Finally, for $p=2$ this quantity can be related to the coefficients of Pauli decomposition of a Hamiltonian going through a mixing operator (see Appendix \ref{sec:2-norm_TPE}).  
We will denote these quantities as $\lambda_1$, $\lambda_2$ and $\lambda_\infty$. 




We estimated the values of $\lambda$ for $t=2$ different gate families by the following numerical procedure. The Haar-averaged tensor product $\mathbb{E}_{Haar} (U^{\otimes t} \otimes (U^*)^{\otimes t})$ is constructed explicitly using exact formulas \cite{mcclean_barren_2018,Poland_Beer_Osborne_2020}. For the two-qubit blocks, we pick the parameters uniformly at random and average the resulting tensor product over $N=500000$ trials. From this average, we estimate $\lambda$.

To estimate the error of this method, we also evaluated $\lambda$ for an ensemble of matrices distributed according to the Haar measure. For $N \rightarrow \infty$, the estimate should converge to zero as $1/\sqrt{N}$. Hence, the numerical value of $\lambda$ shows the typical scale of sampling error. For $N=500000$ trials, we observed the values of $\lambda_1 \approx \lambda_\infty = 0.022$, $\lambda_2 = 0.0028$.

The results of numerical experiments are summarised in Table \ref{tab:local_designs}. As expected, the more sophisticated ans\"atze are usually better at approximating a 2-design. Note further that the block implemented according to the Cartan decomposition of $SU(4)$ \cite{khaneja_cartan_2000,khaneja_time_2001} is not an exact 2-design, although it is capable of preparing any two-qubit gate. Nonetheless, among the gate families studied, this block is the closest approximation of a 2-design.

With a similar numerical experiment for $t=1$, we found that all blocks, except the particle-conserving block \cite{barkoutsos_quantum_2018}, are also exact 1-designs up to sampling tolerance. 

\begin{table}[]
    \centering
    \begin{tabular}{|c|c|c|c|c|}
    \hline
        Block & Circuit diagram or matrix & $\lambda_1$ & $\lambda_\infty$ & $\lambda_2$\\
        \hline
        $X$, $Z$, and $ZZ$ rotations &  
            $\Qcircuit @C=1.0em @R=1.0em {
                   \quad & \gate{R_Z} & \multigate{1}{R_{ZZ}} & \gate{R_X} & \qw \\
                   \quad & \gate{R_Z} & \ghost{R_{ZZ}} & \gate{R_X} & \qw \\
               }$
        & 0.95 & 1.80 & 0.87\\
        \hline 
        Universal gates and a CNOT &
            $\Qcircuit @C=1.0em @R=1.0em {
           \quad & \gate{U_3} & \ctrl{1} & \gate{U_3} & \qw \\
           \quad & \gate{U_3} & \targ & \gate{U_3} & \qw \\
            }$
        & 0.68 & 0.69 & 0.42\\
        \hline
        $Y$ rotations and a CZ \cite{cerezo_cost-function-dependent_2020} &            
            $\Qcircuit @C=1.0em @R=1.0em {
           \quad & \gate{R_Y} & \ctrl{1} & \gate{R_Y} & \qw \\
           \quad & \gate{R_Y} & \ctrl{-1} & \gate{R_Y} & \qw \\
            }$ & 1.76 & 1.76 & 1.00\\
        \hline
        Number-conserving \cite{barkoutsos_quantum_2018} & 
        $
        \begin{pmatrix}
        1 & 0 & 0 & 0 \\
        0 & \cos(\theta_1) & e^{i\theta_2} \sin(\theta_1) & 0 \\
        0 & e^{-i\theta_2} \sin(\theta_1) & -\cos(\theta_1) & 0 \\
        0 & 0 & 0 & 1 \\
        \end{pmatrix}
        $
        & 2.40 & 2.40 & 1.00\\
        \hline
        Cartan decomposition \cite{khaneja_cartan_2000,khaneja_time_2001} &
        $\Qcircuit @C=1.0em @R=1.0em {
       \quad & \gate{U_3} & \multigate{1}{R_{XX}} & \multigate{1}{R_{YY}} & \multigate{1}{R_{ZZ}}& \gate{U_3} & \qw \\
       \quad & \gate{U_3} & \ghost{R_{XX}} & \ghost{R_{YY}} & \ghost{R_{ZZ}} & \gate{U_3} & \qw \\
        }$
            & 0.25 & 0.25 & 0.17\\
    \hline
    \end{tabular}
    \caption{Proximity to the 2-tensor product expander for different two-qubit blocks, estimated by random sampling.}
    \label{tab:local_designs}
\end{table}


\subsection{Plateau dependence}
\label{subsec:plateau_numeric}

\begin{figure}
    \centering
    \begin{subfigure}{.48\linewidth}
        \centering
        \includegraphics[width=\textwidth]{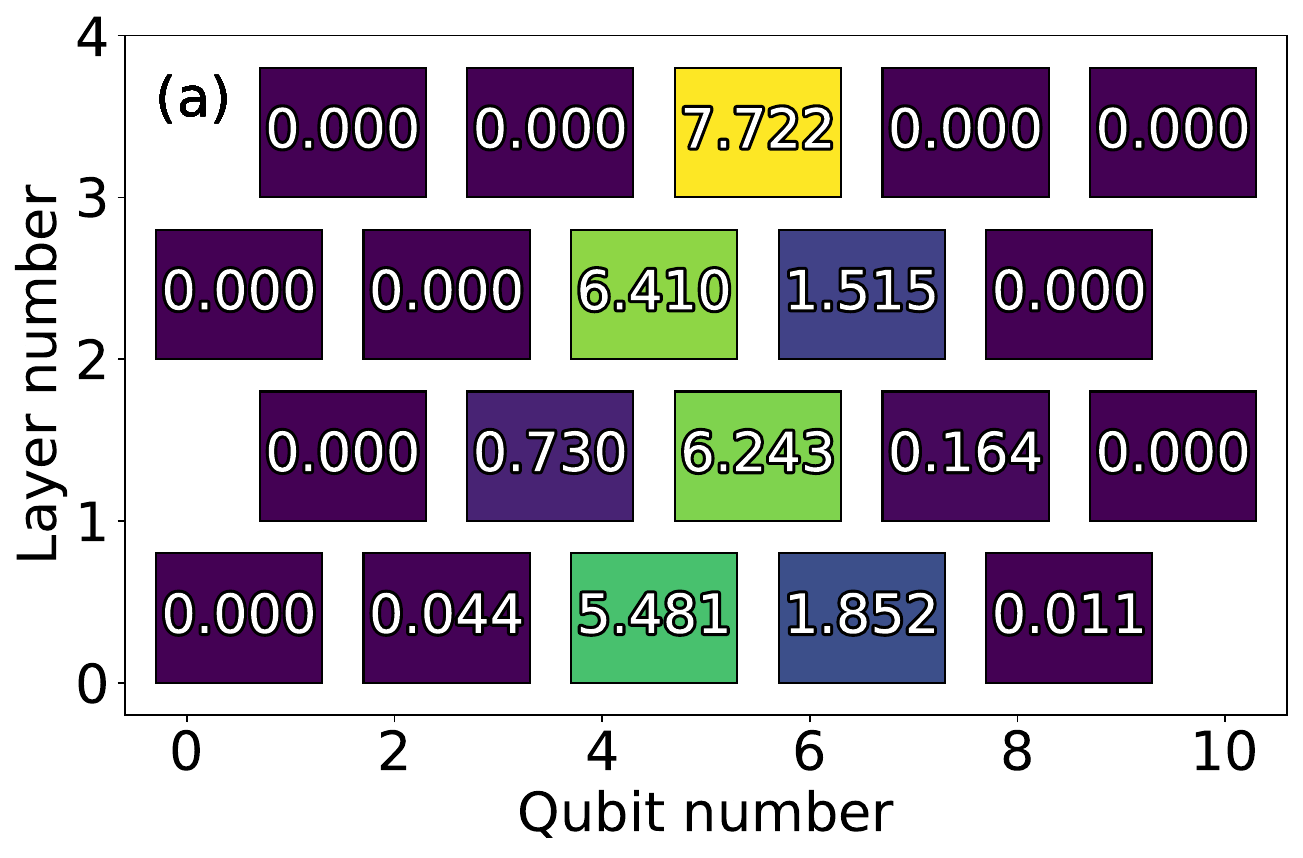}
    \end{subfigure}\begin{subfigure}{.48\linewidth}
        \centering
        \includegraphics[width=\textwidth]{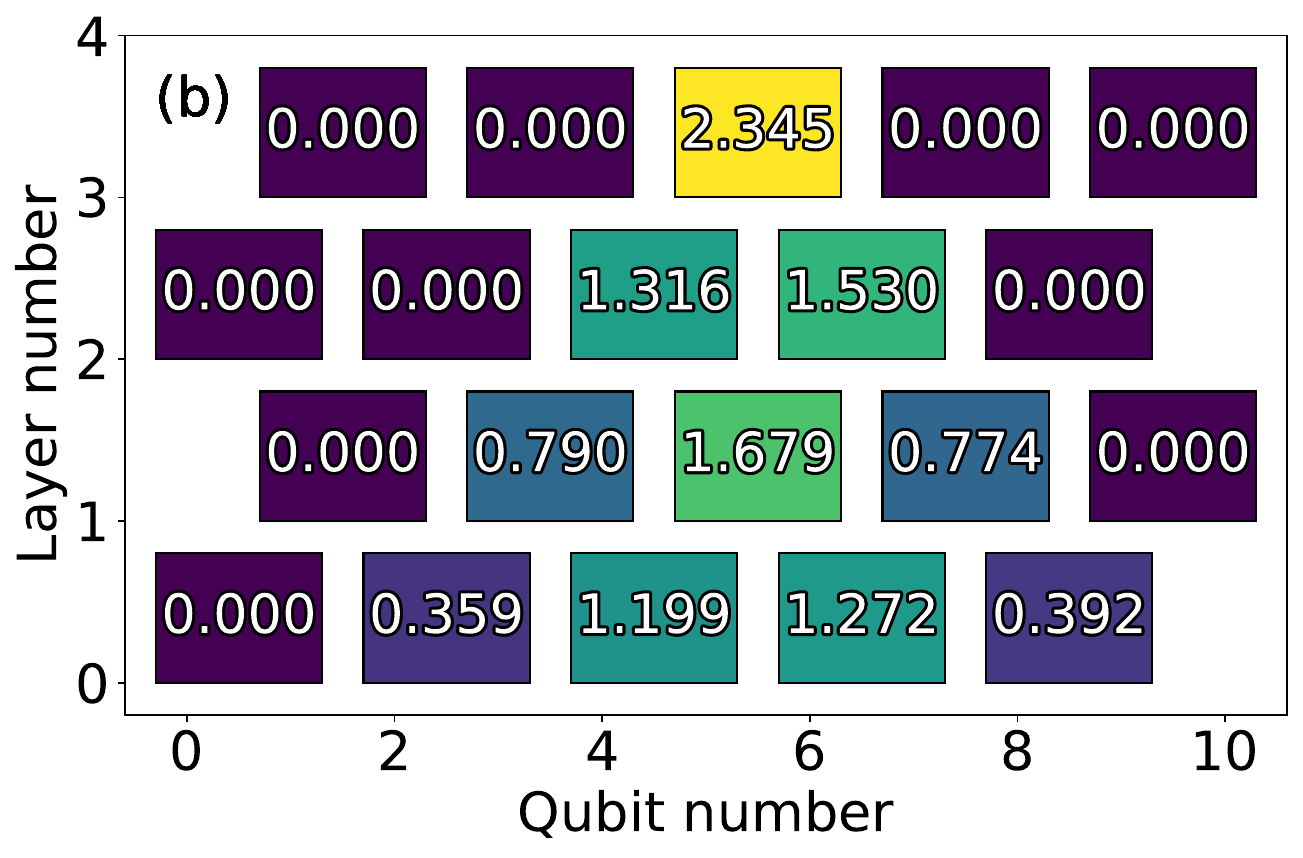}
    \end{subfigure}
    \begin{subfigure}{.48\linewidth}
        \centering
        \includegraphics[width=\textwidth]{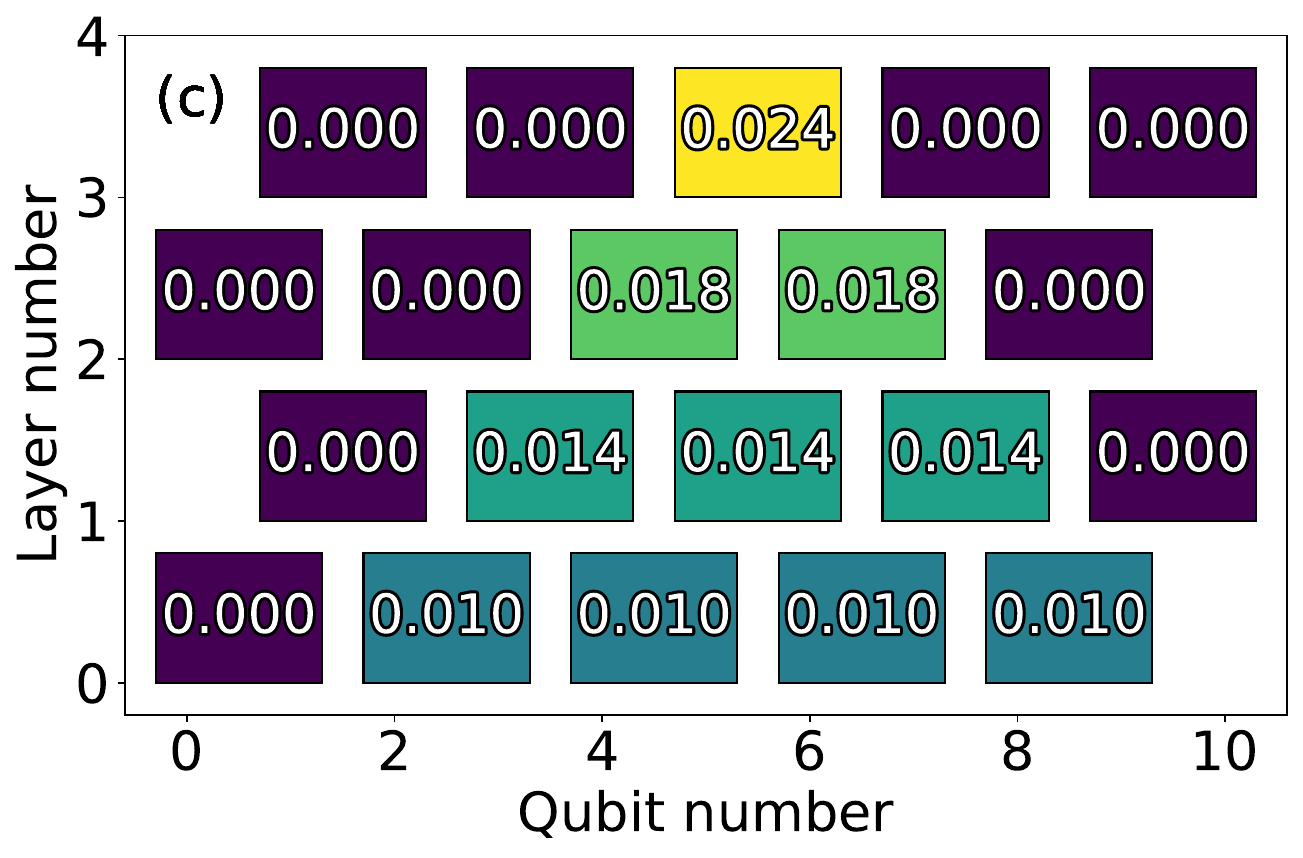}
    \end{subfigure}
    \caption{Derivative variances for $H = X_5$, averaged over parameters in each ansatz block. The numbers in the boxes denote $\Var \partial_\theta E \cdot 100$. Qubit number 10 is identified with qubit number 0. (a) Numerical result for an ansatz with blocks of $X,Z,ZZ$ rotations. (b) Numerical result for blocks implemented according to the Cartan decomposition. (c) Lower bound given in Theorem \ref{thm:main}.}
    \label{fig:one-local}
\end{figure}

To estimate gradients, we used the following analytical procedure \cite{mitarai_quantum_2018,schuld_evaluating_2019}: let $f(\theta)$ be the cost function, and $\theta$ a parameter which is included in the quantum circuit in a gate like $\exp(\rmi F\theta / 2)$ for some Pauli operator $F$. Then the derivative w.r.t.\ this parameter is equal to $(f(\theta + \pi /2) - f(\theta - \pi / 2))/2$. The simulations assume noise-free conditions and use the statevector simulator provided by Qiskit.

\begin{figure}
    \centering
    \begin{subfigure}{.48\linewidth}
        \centering
        \includegraphics[width=\linewidth]{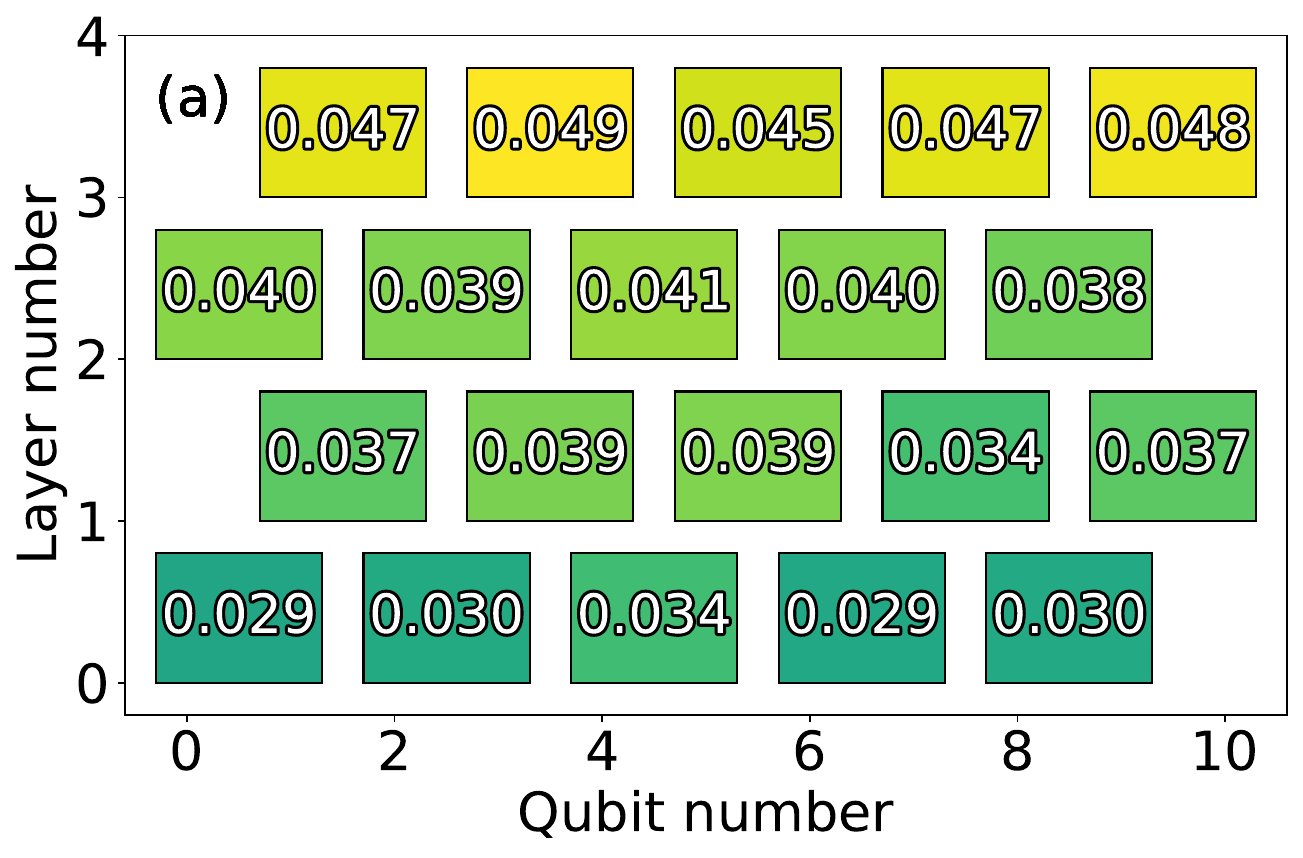}
    \end{subfigure}
    \begin{subfigure}{.48\linewidth}
        \centering
        \includegraphics[width=\linewidth]{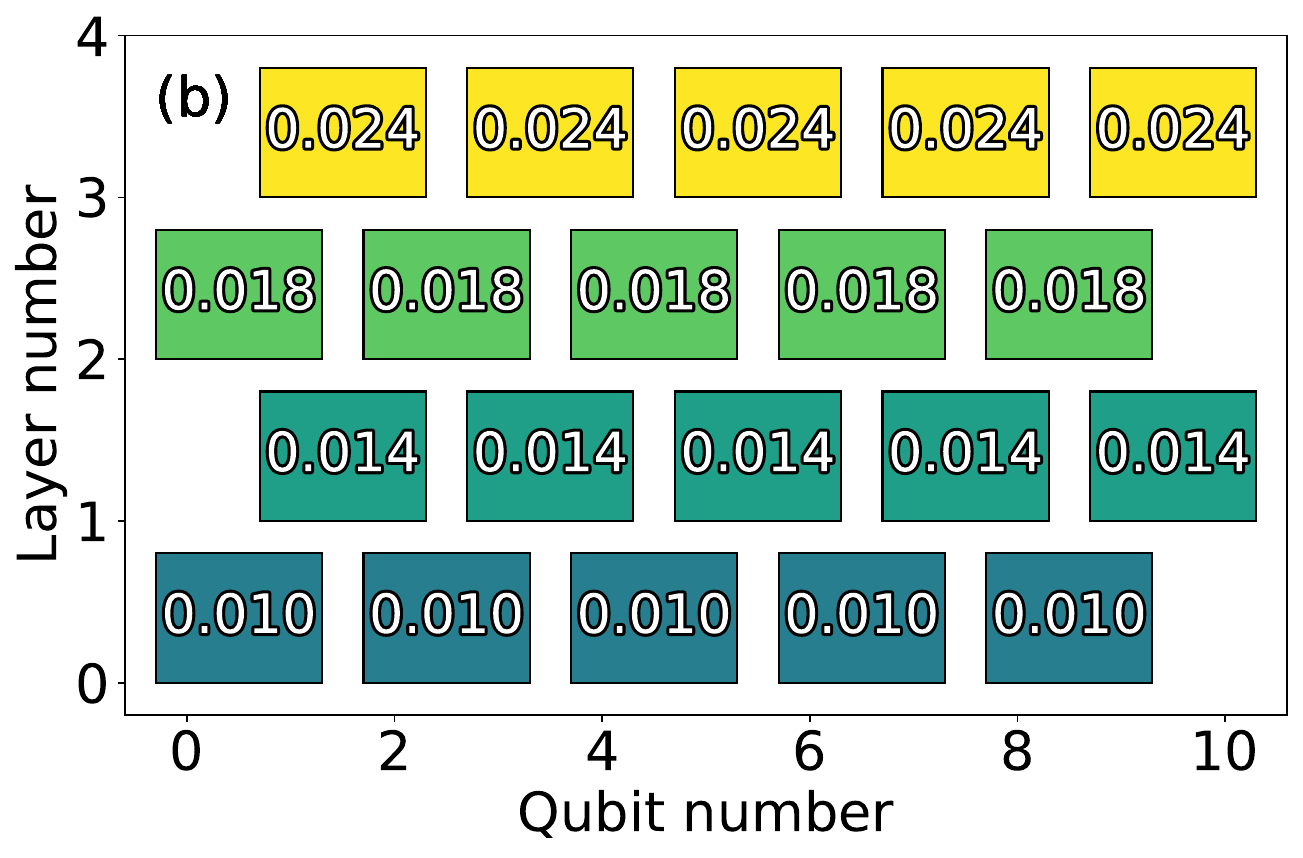}
    \end{subfigure}
    \caption{Derivative variances for $H = X^{\otimes n}$, numerical estimate (a) and the lower bound (b).}
    \label{fig:n-local}
\end{figure}

The first Hamiltonian we tested our predictions on is the single-qubit Hamiltonian $H = X_5$ acting on $n=10$ qubits. The qubits are enumerated starting from zero. 
For $N = 400$ samples, the derivative with respect to each parameter was evaluated, then the variances were averaged over each block. We performed two numerical experiments with different two-qubit blocks from Table \ref{tab:local_designs}: one with blocks of $X$, $Z$, and $ZZ$ rotations, and the other with blocks implemented according to the Cartan decomposition of $SU(4)$. The ansatz used was the checkerboard ansatz with ring connectivity. The result of the numerical test is shown in Fig.~\ref{fig:one-local}. 

\begin{figure}
    \centering
    \includegraphics[width=\linewidth]{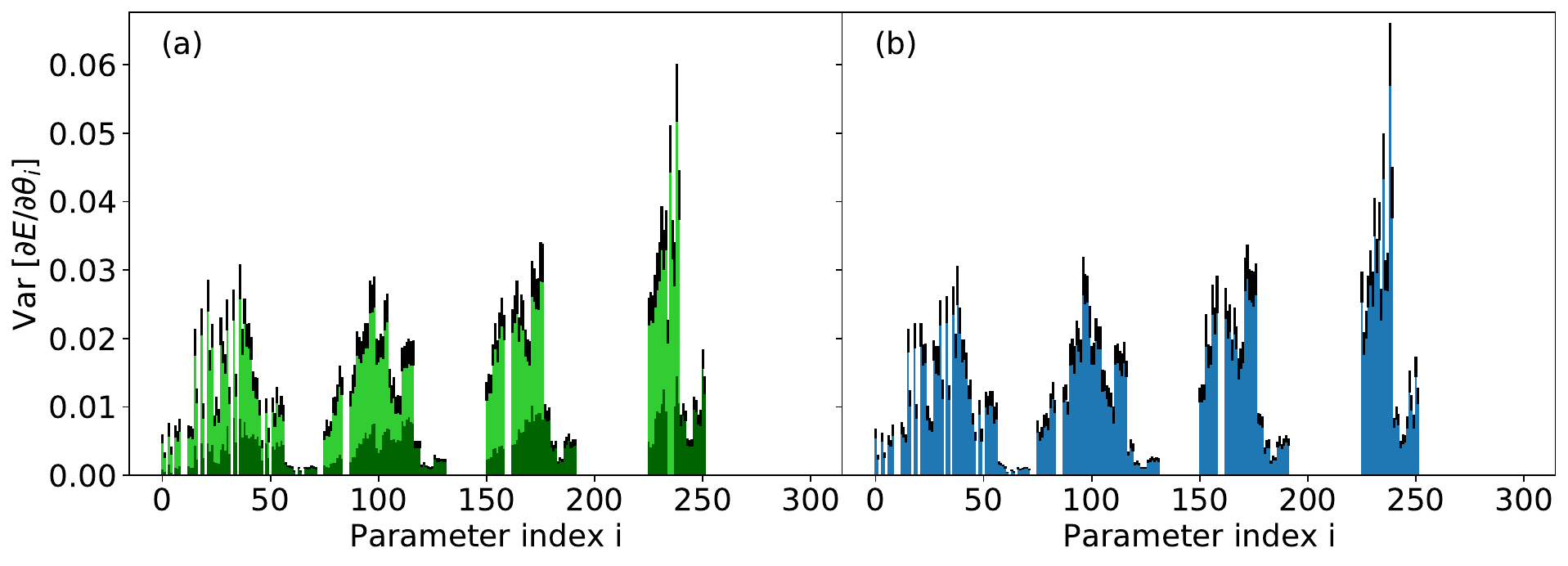}
    \caption{Variances of the cost function derivatives with respect to different ansatz parameters for $H_1 = X_5 X_6$, $H_2 = X_4 X_5$ (a), and their sum $H_1 + H_2$ (b).}
    \label{fig:additive}
\end{figure}

In both experiments, the causal cone structure is evident, as is evident the tendency of the gradients to decrease with the decreasing number of layer. However, the Cartan decomposed blocks show smoother results. This result is consistent with the fact that the first block type is further from a 2-design. The condition that the block can be further decomposed into two independent local 2-designs is also violated in the first case. Because of these factors, the gradients are uneven.

The theoretical lower bound is fulfilled by a large margin in both cases. The lower bound also does not catch the difference of gradients within one layer, which tend to be more significant in the middle of the causal cone as opposed to the edges of the cone, where the gradients are much smaller.

\begin{figure}
    \centering
    \includegraphics[width=0.7\linewidth]{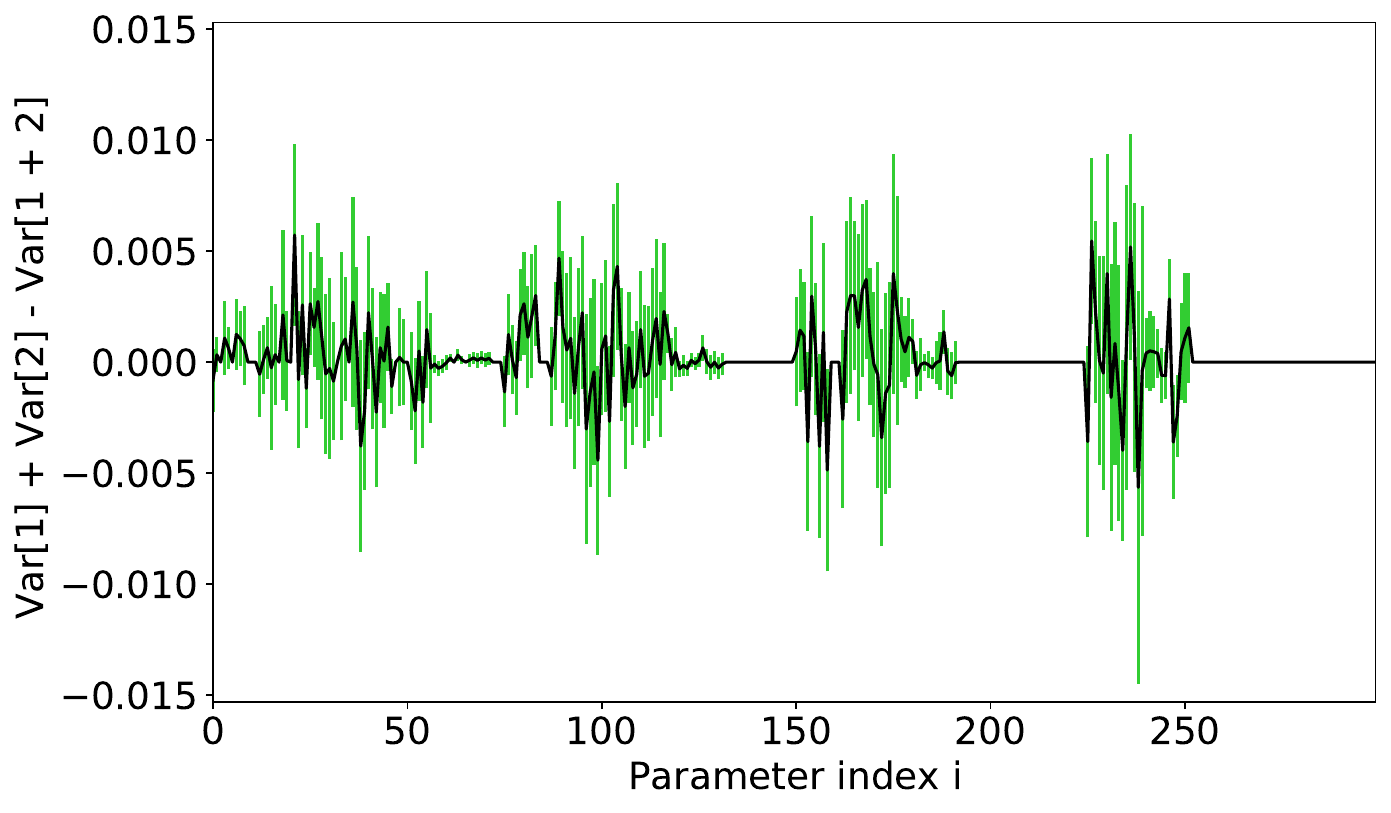}
    \caption{Difference between the variances plotted in Fig.~\ref{fig:additive}. Error bars denote one standard error.}
    \label{fig:additive_delta}
\end{figure}

Figure \ref{fig:n-local} shows the results of a similar numerical test for $H = X^{\otimes n}$ for $n = 10$. Here, the ``Cartan decomposition'' blocks were used in the ansatz. As implied by the lower bound and in accordance with the results previously found in the literature \cite{cerezo_cost-function-dependent_2020}, this $n$-local Hamiltonian exhibits barren plateaus even for a very shallow ansatz.

According to \eqref{eq:paulis_decouple}, variances for a Hamiltonian consisting of several Pauli strings are equal to the sum of variances computed for each Pauli string independently. We tested that prediction on a pair of Hamiltonians $H_1 = X_4 X_5$, $H_2 = X_5 X_6$. In this test, the ansatz acts on 10 qubits and consists of 4 layers of ``Cartan decomposition'' blocks. The variances for $H_1$ and $H_2$ separately are shown as a stacked bar chart in Fig.~\ref{fig:additive}a. Each bar corresponds to a parameter $\theta_i$ in the ansatz. Fig.~\ref{fig:additive}b shows the variances for $H_1 + H_2$. The qualitative agreement between the graphs is evident, and the differences for each parameter of the ansatz (shown in Fig.~\ref{fig:additive_delta}) are close to zero, up to the standard errors of the samples.

\subsection{Alternative ansatz architectures}
\label{subsec:alt_ansatz}

The width of the causal cone depends on the ansatz structure. Conversely, some ansatz structures may be less prone to barren plateaus. We performed the same numerical tests for two more circuit architectures that are better suited for NISQ devices.

\subsubsection{Checkerboard with open boundary conditions}

\begin{figure}
    \centering
    \begin{subfigure}{.48\linewidth}
        \centering
        \includegraphics[width=\linewidth]{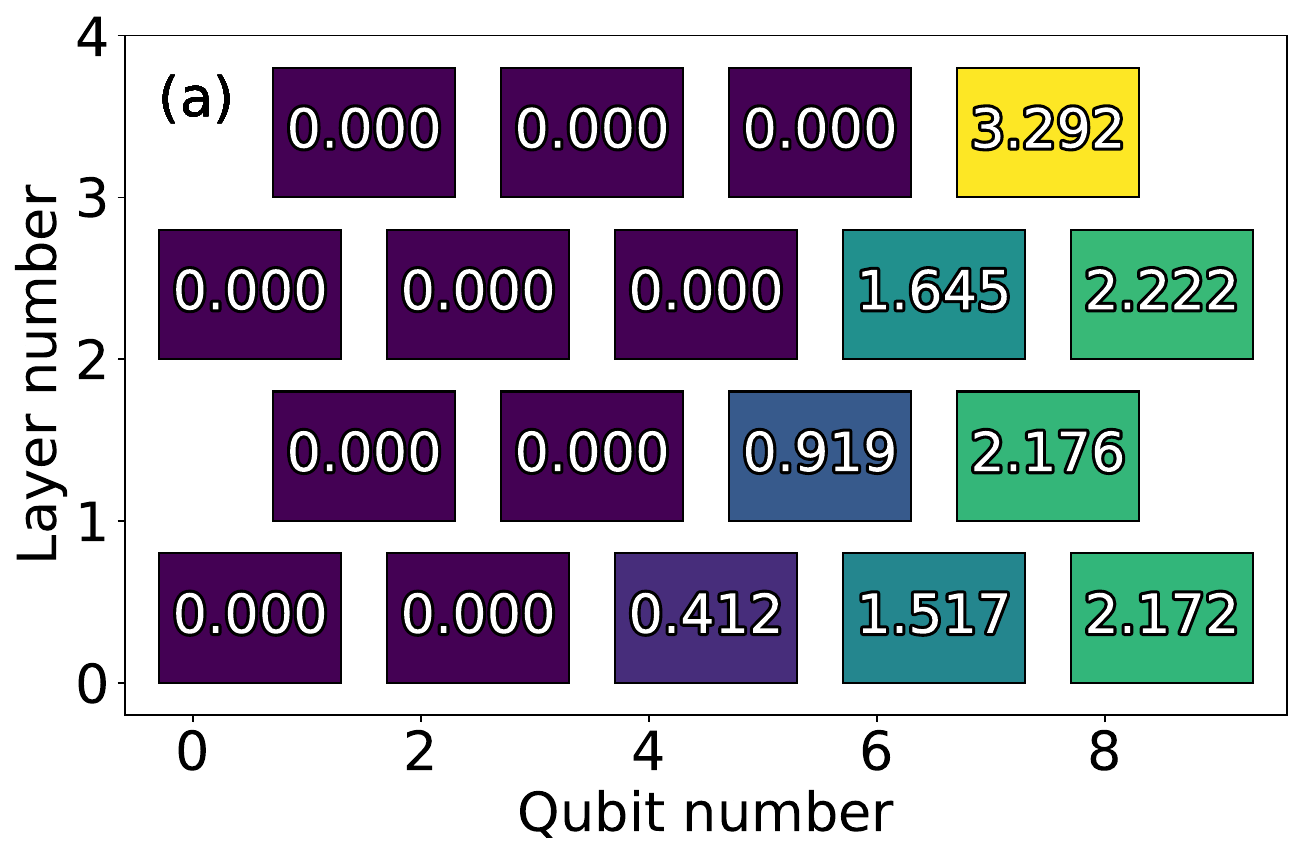}
    \end{subfigure}
    \begin{subfigure}{.48\linewidth}
        \centering
        \includegraphics[width=\linewidth]{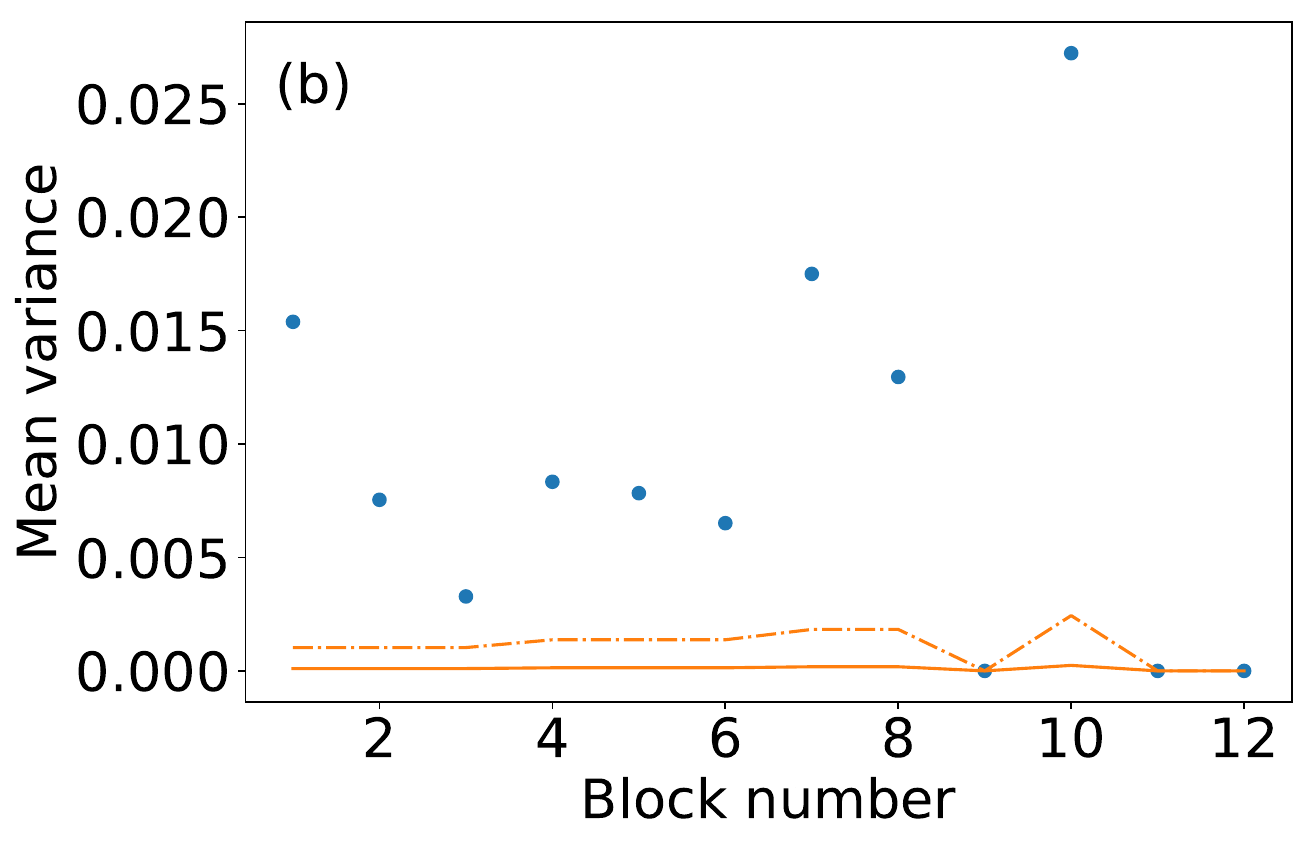}
    \end{subfigure}
    \caption{Blockwise averaged values of derivative variances with respect to one-local Pauli strings. (a) Numerical result for a line-connected checkerboard ansatz. (b) Numerical result for a two-dimensional lattice ansatz. Dots: numerical values, line: lower bound,  dot-dashed line: lower bound multiplied by 10.}
    \label{fig:alt_connect}
\end{figure}

For certain quantum computing platforms, e.g. Calcium ions and Rydberg atoms, it is easiest to arrange qubits in a line and perform entangling gates acting on adjacent qubits. Unlike ring connectivity, this structure does not use direct coupling of the first qubit with the last qubit. Thus, the qubits closer to the edge will have narrower causal cones, and possibly higher values of the gradients. Fig.~\ref{fig:alt_connect}a shows the behavior of derivatives for such an architecture, for $H = X_8$. In comparison with the ring connectivity (Fig. \ref{fig:one-local}b), the gradient variances are significantly larger.

\subsubsection{Two-dimensional lattice}

We also tested the predictions of Theorem \ref{thm:main} on a two-dimensional $3 \times 3$ lattice. The numerical values of the derivative variances, as well as their lower bounds, are depicted in Fig.~\ref{fig:alt_connect}b. The causal cone structure for this ansatz is more convoluted, but it is possible to tell that some ansatz blocks are not included in the causal cone, and hence, the derivative over their parameters is equal to zero. The construction of the two-dimensional lattice ansatz is deferred to Appendix \ref{sec:2d_ansatz}.

\section{Conclusions and future work}
\label{sec:conclusions}

Given a Hamiltonian, we can now estimate its susceptibility to the barren plateaus. One can hence preprocess Hamiltonians in order to make the optimization more viable. For example, a method similar to that of Ref. \cite{ryabinkin_iterative_2019} could be employed.

Our results indicate that the severity of plateaus also depends on the structure of the ansatz. This may mean that some hardware topologies are more suited for VQE than others. For example, our numerical tests demonstrate that in line connectivity of the qubits, the gates on the edges are potentially less prone to vanishing derivatives than those in the middle.

The numerical tests provided in this paper can be implemented in hardware as well. The computational cost of simulating a quantum computer using the best known methods is exponential either in time or in memory, while estimation of the gradients in quantum hardware (up to fixed absolute tolerance) is linear in the number of ansatz parameters, and possibly even sub-linear in the cardinality of the problem Hamiltonian, if clever simultaneous measurement strategies are used \cite{verteletskyi_measurement_2019}. 

An interesting direction of research is to study the efficiency of techniques that were already proposed or used for circumventing the barren plateau problem, like optimizing only a few parameters simultaneously \cite{grant_initialization_2019,Skolik_2020}, using adiabatically assisted VQE \cite{garcia-saez_addressing_2018,bravo-prieto_scaling_2020}, or dynamically updating the ansatz structure \cite{Bilkis_Cerezo_Verdon_Coles_Cincio_2021,grimsley_adaptive_2019}.


\section*{Acknowledgments}
AU acknowledges support from the Russian Foundation for Basic Research under project No.~19-31-90159. JB acknowledges support from the Leading Research Center on Quantum Computing (Agreement No.~014/20). 



\bibliography{references.bib}

\begin{thebibliography}{49}%
\makeatletter
\providecommand \@ifxundefined [1]{%
 \@ifx{#1\undefined}
}%
\providecommand \@ifnum [1]{%
 \ifnum #1\expandafter \@firstoftwo
 \else \expandafter \@secondoftwo
 \fi
}%
\providecommand \@ifx [1]{%
 \ifx #1\expandafter \@firstoftwo
 \else \expandafter \@secondoftwo
 \fi
}%
\providecommand \natexlab [1]{#1}%
\providecommand \enquote  [1]{``#1''}%
\providecommand \bibnamefont  [1]{#1}%
\providecommand \bibfnamefont [1]{#1}%
\providecommand \citenamefont [1]{#1}%
\providecommand \href@noop [0]{\@secondoftwo}%
\providecommand \href [0]{\begingroup \@sanitize@url \@href}%
\providecommand \@href[1]{\@@startlink{#1}\@@href}%
\providecommand \@@href[1]{\endgroup#1\@@endlink}%
\providecommand \@sanitize@url [0]{\catcode `\\12\catcode `\$12\catcode
  `\&12\catcode `\#12\catcode `\^12\catcode `\_12\catcode `\%12\relax}%
\providecommand \@@startlink[1]{}%
\providecommand \@@endlink[0]{}%
\providecommand \url  [0]{\begingroup\@sanitize@url \@url }%
\providecommand \@url [1]{\endgroup\@href {#1}{\urlprefix }}%
\providecommand \urlprefix  [0]{URL }%
\providecommand \Eprint [0]{\href }%
\providecommand \doibase [0]{http://dx.doi.org/}%
\providecommand \selectlanguage [0]{\@gobble}%
\providecommand \bibinfo  [0]{\@secondoftwo}%
\providecommand \bibfield  [0]{\@secondoftwo}%
\providecommand \translation [1]{[#1]}%
\providecommand \BibitemOpen [0]{}%
\providecommand \bibitemStop [0]{}%
\providecommand \bibitemNoStop [0]{.\EOS\space}%
\providecommand \EOS [0]{\spacefactor3000\relax}%
\providecommand \BibitemShut  [1]{\csname bibitem#1\endcsname}%
\let\auto@bib@innerbib\@empty
\bibitem [{\citenamefont {McClean}\ \emph {et~al.}(2016)\citenamefont
  {McClean}, \citenamefont {Romero}, \citenamefont {Babbush},\ and\
  \citenamefont {Aspuru-Guzik}}]{mcclean_theory_2016}%
  \BibitemOpen
  \bibfield  {author} {\bibinfo {author} {\bibfnamefont {J.~R.}\ \bibnamefont
  {McClean}}, \bibinfo {author} {\bibfnamefont {J.}~\bibnamefont {Romero}},
  \bibinfo {author} {\bibfnamefont {R.}~\bibnamefont {Babbush}}, \ and\
  \bibinfo {author} {\bibfnamefont {A.}~\bibnamefont {Aspuru-Guzik}},\
  }\bibfield  {title} {\enquote {\bibinfo {title} {The theory of variational
  hybrid quantum-classical algorithms},}\ }\href {\doibase
  10.1088/1367-2630/18/2/023023} {\bibfield  {journal} {\bibinfo  {journal}
  {New Journal of Physics}\ }\textbf {\bibinfo {volume} {18}},\ \bibinfo
  {pages} {023023} (\bibinfo {year} {2016})},\ \bibinfo {note} {arXiv:
  1509.04279}\BibitemShut {NoStop}%
\bibitem [{\citenamefont {Peruzzo}\ \emph {et~al.}(2014)\citenamefont
  {Peruzzo}, \citenamefont {McClean}, \citenamefont {Shadbolt}, \citenamefont
  {Yung}, \citenamefont {Zhou}, \citenamefont {Love}, \citenamefont
  {Aspuru-Guzik},\ and\ \citenamefont
  {O{\textquoteright}Brien}}]{peruzzo_variational_2014}%
  \BibitemOpen
  \bibfield  {author} {\bibinfo {author} {\bibfnamefont {A.}~\bibnamefont
  {Peruzzo}}, \bibinfo {author} {\bibfnamefont {J.}~\bibnamefont {McClean}},
  \bibinfo {author} {\bibfnamefont {P.}~\bibnamefont {Shadbolt}}, \bibinfo
  {author} {\bibfnamefont {M.-H.}\ \bibnamefont {Yung}}, \bibinfo {author}
  {\bibfnamefont {X.-Q.}\ \bibnamefont {Zhou}}, \bibinfo {author}
  {\bibfnamefont {P.~J.}\ \bibnamefont {Love}}, \bibinfo {author}
  {\bibfnamefont {A.}~\bibnamefont {Aspuru-Guzik}}, \ and\ \bibinfo {author}
  {\bibfnamefont {J.~L.}\ \bibnamefont {O{\textquoteright}Brien}},\ }\bibfield
  {title} {\enquote {\bibinfo {title} {A variational eigenvalue solver on a
  photonic quantum processor},}\ }\href {\doibase 10.1038/ncomms5213}
  {\bibfield  {journal} {\bibinfo  {journal} {Nature Communications}\ }\textbf
  {\bibinfo {volume} {5}},\ \bibinfo {pages} {4213} (\bibinfo {year}
  {2014})}\BibitemShut {NoStop}%
\bibitem [{\citenamefont {Kandala}\ \emph {et~al.}(2017)\citenamefont
  {Kandala}, \citenamefont {Mezzacapo}, \citenamefont {Temme}, \citenamefont
  {Takita}, \citenamefont {Brink}, \citenamefont {Chow},\ and\ \citenamefont
  {Gambetta}}]{kandala_hardware-efficient_2017}%
  \BibitemOpen
  \bibfield  {author} {\bibinfo {author} {\bibfnamefont {A.}~\bibnamefont
  {Kandala}}, \bibinfo {author} {\bibfnamefont {A.}~\bibnamefont {Mezzacapo}},
  \bibinfo {author} {\bibfnamefont {K.}~\bibnamefont {Temme}}, \bibinfo
  {author} {\bibfnamefont {M.}~\bibnamefont {Takita}}, \bibinfo {author}
  {\bibfnamefont {M.}~\bibnamefont {Brink}}, \bibinfo {author} {\bibfnamefont
  {J.~M.}\ \bibnamefont {Chow}}, \ and\ \bibinfo {author} {\bibfnamefont
  {J.~M.}\ \bibnamefont {Gambetta}},\ }\bibfield  {title} {\enquote {\bibinfo
  {title} {Hardware-efficient {Variational} {Quantum} {Eigensolver} for {Small}
  {Molecules} and {Quantum} {Magnets}},}\ }\href {\doibase 10.1038/nature23879}
  {\bibfield  {journal} {\bibinfo  {journal} {Nature}\ }\textbf {\bibinfo
  {volume} {549}},\ \bibinfo {pages} {242--246} (\bibinfo {year} {2017})},\
  \bibinfo {note} {arXiv: 1704.05018}\BibitemShut {NoStop}%
\bibitem [{\citenamefont {Barkoutsos}\ \emph {et~al.}(2018)\citenamefont
  {Barkoutsos}, \citenamefont {Gonthier}, \citenamefont {Sokolov},
  \citenamefont {Moll}, \citenamefont {Salis}, \citenamefont {Fuhrer},
  \citenamefont {Ganzhorn}, \citenamefont {Egger}, \citenamefont {Troyer},
  \citenamefont {Mezzacapo}, \citenamefont {Filipp},\ and\ \citenamefont
  {Tavernelli}}]{barkoutsos_quantum_2018}%
  \BibitemOpen
  \bibfield  {author} {\bibinfo {author} {\bibfnamefont {P.~K.}\ \bibnamefont
  {Barkoutsos}}, \bibinfo {author} {\bibfnamefont {J.~F.}\ \bibnamefont
  {Gonthier}}, \bibinfo {author} {\bibfnamefont {I.}~\bibnamefont {Sokolov}},
  \bibinfo {author} {\bibfnamefont {N.}~\bibnamefont {Moll}}, \bibinfo {author}
  {\bibfnamefont {G.}~\bibnamefont {Salis}}, \bibinfo {author} {\bibfnamefont
  {A.}~\bibnamefont {Fuhrer}}, \bibinfo {author} {\bibfnamefont
  {M.}~\bibnamefont {Ganzhorn}}, \bibinfo {author} {\bibfnamefont {D.~J.}\
  \bibnamefont {Egger}}, \bibinfo {author} {\bibfnamefont {M.}~\bibnamefont
  {Troyer}}, \bibinfo {author} {\bibfnamefont {A.}~\bibnamefont {Mezzacapo}},
  \bibinfo {author} {\bibfnamefont {S.}~\bibnamefont {Filipp}}, \ and\ \bibinfo
  {author} {\bibfnamefont {I.}~\bibnamefont {Tavernelli}},\ }\bibfield  {title}
  {\enquote {\bibinfo {title} {Quantum algorithms for electronic structure
  calculations: particle/hole {Hamiltonian} and optimized wavefunction
  expansions},}\ }\href {\doibase 10.1103/PhysRevA.98.022322} {\bibfield
  {journal} {\bibinfo  {journal} {Physical Review A}\ }\textbf {\bibinfo
  {volume} {98}} (\bibinfo {year} {2018}),\ 10.1103/PhysRevA.98.022322},\
  \bibinfo {note} {arXiv: 1805.04340}\BibitemShut {NoStop}%
\bibitem [{\citenamefont {Cade}\ \emph {et~al.}(2019)\citenamefont {Cade},
  \citenamefont {Mineh}, \citenamefont {Montanaro},\ and\ \citenamefont
  {Stanisic}}]{cade_strategies_2019}%
  \BibitemOpen
  \bibfield  {author} {\bibinfo {author} {\bibfnamefont {C.}~\bibnamefont
  {Cade}}, \bibinfo {author} {\bibfnamefont {L.}~\bibnamefont {Mineh}},
  \bibinfo {author} {\bibfnamefont {A.}~\bibnamefont {Montanaro}}, \ and\
  \bibinfo {author} {\bibfnamefont {S.}~\bibnamefont {Stanisic}},\ }\bibfield
  {title} {\enquote {\bibinfo {title} {Strategies for solving the
  {Fermi}-{Hubbard} model on near-term quantum computers},}\ }\href
  {http://arxiv.org/abs/1912.06007} {\bibfield  {journal} {\bibinfo  {journal}
  {arXiv:1912.06007 [quant-ph]}\ } (\bibinfo {year} {2019})},\ \bibinfo {note}
  {arXiv: 1912.06007}\BibitemShut {NoStop}%
\bibitem [{\citenamefont {Farhi}, \citenamefont {Goldstone},\ and\
  \citenamefont {Gutmann}(2014)}]{farhi_quantum_2014}%
  \BibitemOpen
  \bibfield  {author} {\bibinfo {author} {\bibfnamefont {E.}~\bibnamefont
  {Farhi}}, \bibinfo {author} {\bibfnamefont {J.}~\bibnamefont {Goldstone}}, \
  and\ \bibinfo {author} {\bibfnamefont {S.}~\bibnamefont {Gutmann}},\
  }\bibfield  {title} {\enquote {\bibinfo {title} {A {Quantum} {Approximate}
  {Optimization} {Algorithm}},}\ }\href {http://arxiv.org/abs/1411.4028}
  {\bibfield  {journal} {\bibinfo  {journal} {arXiv:1411.4028 [quant-ph]}\ }
  (\bibinfo {year} {2014})},\ \bibinfo {note} {arXiv: 1411.4028}\BibitemShut
  {NoStop}%
\bibitem [{\citenamefont {Willsch}\ \emph {et~al.}(2019)\citenamefont
  {Willsch}, \citenamefont {Willsch}, \citenamefont {Jin}, \citenamefont
  {De~Raedt},\ and\ \citenamefont {Michielsen}}]{willsch_benchmarking_2019}%
  \BibitemOpen
  \bibfield  {author} {\bibinfo {author} {\bibfnamefont {M.}~\bibnamefont
  {Willsch}}, \bibinfo {author} {\bibfnamefont {D.}~\bibnamefont {Willsch}},
  \bibinfo {author} {\bibfnamefont {F.}~\bibnamefont {Jin}}, \bibinfo {author}
  {\bibfnamefont {H.}~\bibnamefont {De~Raedt}}, \ and\ \bibinfo {author}
  {\bibfnamefont {K.}~\bibnamefont {Michielsen}},\ }\bibfield  {title}
  {\enquote {\bibinfo {title} {Benchmarking the {Quantum} {Approximate}
  {Optimization} {Algorithm}},}\ }\href {http://arxiv.org/abs/1907.02359}
  {\bibfield  {journal} {\bibinfo  {journal} {arXiv:1907.02359 [quant-ph]}\ }
  (\bibinfo {year} {2019})},\ \bibinfo {note} {arXiv: 1907.02359}\BibitemShut
  {NoStop}%
\bibitem [{\citenamefont {Morales}, \citenamefont {Tlyachev},\ and\
  \citenamefont {Biamonte}(2018)}]{morales_variational_2018}%
  \BibitemOpen
  \bibfield  {author} {\bibinfo {author} {\bibfnamefont {M.~E.~S.}\
  \bibnamefont {Morales}}, \bibinfo {author} {\bibfnamefont {T.}~\bibnamefont
  {Tlyachev}}, \ and\ \bibinfo {author} {\bibfnamefont {J.}~\bibnamefont
  {Biamonte}},\ }\bibfield  {title} {\enquote {\bibinfo {title} {Variational
  learning of {Grover}'s quantum search algorithm},}\ }\href {\doibase
  10.1103/PhysRevA.98.062333} {\bibfield  {journal} {\bibinfo  {journal}
  {Physical Review A}\ }\textbf {\bibinfo {volume} {98}} (\bibinfo {year}
  {2018}),\ 10.1103/PhysRevA.98.062333}\BibitemShut {NoStop}%
\bibitem [{\citenamefont {Akshay}\ \emph {et~al.}(2020)\citenamefont {Akshay},
  \citenamefont {Philathong}, \citenamefont {Morales},\ and\ \citenamefont
  {Biamonte}}]{Akshay_Philathong_Morales_Biamonte_2020}%
  \BibitemOpen
  \bibfield  {author} {\bibinfo {author} {\bibfnamefont {V.}~\bibnamefont
  {Akshay}}, \bibinfo {author} {\bibfnamefont {H.}~\bibnamefont {Philathong}},
  \bibinfo {author} {\bibfnamefont {M.~E.~S.}\ \bibnamefont {Morales}}, \ and\
  \bibinfo {author} {\bibfnamefont {J.~D.}\ \bibnamefont {Biamonte}},\
  }\bibfield  {title} {\enquote {\bibinfo {title} {Reachability deficits in
  quantum approximate optimization},}\ }\href {\doibase
  10.1103/PhysRevLett.124.090504} {\bibfield  {journal} {\bibinfo  {journal}
  {Physical Review Letters}\ }\textbf {\bibinfo {volume} {124}},\ \bibinfo
  {pages} {090504} (\bibinfo {year} {2020})}\BibitemShut {NoStop}%
\bibitem [{\citenamefont {Romero}, \citenamefont {Olson},\ and\ \citenamefont
  {Aspuru-Guzik}(2017)}]{romero_quantum_2017}%
  \BibitemOpen
  \bibfield  {author} {\bibinfo {author} {\bibfnamefont {J.}~\bibnamefont
  {Romero}}, \bibinfo {author} {\bibfnamefont {J.~P.}\ \bibnamefont {Olson}}, \
  and\ \bibinfo {author} {\bibfnamefont {A.}~\bibnamefont {Aspuru-Guzik}},\
  }\bibfield  {title} {\enquote {\bibinfo {title} {Quantum autoencoders for
  efficient compression of quantum data},}\ }\href {\doibase
  10.1088/2058-9565/aa8072} {\bibfield  {journal} {\bibinfo  {journal} {Quantum
  Science and Technology}\ }\textbf {\bibinfo {volume} {2}},\ \bibinfo {pages}
  {045001} (\bibinfo {year} {2017})}\BibitemShut {NoStop}%
\bibitem [{\citenamefont {Havl{\'i}{\v c}ek}\ \emph {et~al.}(2019)\citenamefont
  {Havl{\'i}{\v c}ek}, \citenamefont {C{\'o}rcoles}, \citenamefont {Temme},
  \citenamefont {Harrow}, \citenamefont {Kandala}, \citenamefont {Chow},\ and\
  \citenamefont {Gambetta}}]{havlicek_supervised_2019}%
  \BibitemOpen
  \bibfield  {author} {\bibinfo {author} {\bibfnamefont {V.}~\bibnamefont
  {Havl{\'i}{\v c}ek}}, \bibinfo {author} {\bibfnamefont {A.~D.}\ \bibnamefont
  {C{\'o}rcoles}}, \bibinfo {author} {\bibfnamefont {K.}~\bibnamefont {Temme}},
  \bibinfo {author} {\bibfnamefont {A.~W.}\ \bibnamefont {Harrow}}, \bibinfo
  {author} {\bibfnamefont {A.}~\bibnamefont {Kandala}}, \bibinfo {author}
  {\bibfnamefont {J.~M.}\ \bibnamefont {Chow}}, \ and\ \bibinfo {author}
  {\bibfnamefont {J.~M.}\ \bibnamefont {Gambetta}},\ }\bibfield  {title}
  {\enquote {\bibinfo {title} {Supervised learning with quantum-enhanced
  feature spaces},}\ }\href {\doibase 10.1038/s41586-019-0980-2} {\bibfield
  {journal} {\bibinfo  {journal} {Nature}\ }\textbf {\bibinfo {volume} {567}},\
  \bibinfo {pages} {209--212} (\bibinfo {year} {2019})}\BibitemShut {NoStop}%
\bibitem [{\citenamefont {Schuld}\ \emph {et~al.}(2020)\citenamefont {Schuld},
  \citenamefont {Bocharov}, \citenamefont {Svore},\ and\ \citenamefont
  {Wiebe}}]{schuld_circuit-centric_2018}%
  \BibitemOpen
  \bibfield  {author} {\bibinfo {author} {\bibfnamefont {M.}~\bibnamefont
  {Schuld}}, \bibinfo {author} {\bibfnamefont {A.}~\bibnamefont {Bocharov}},
  \bibinfo {author} {\bibfnamefont {K.~M.}\ \bibnamefont {Svore}}, \ and\
  \bibinfo {author} {\bibfnamefont {N.}~\bibnamefont {Wiebe}},\ }\bibfield
  {title} {\enquote {\bibinfo {title} {Circuit-centric quantum classifiers},}\
  }\href {\doibase 10.1103/PhysRevA.101.032308} {\bibfield  {journal} {\bibinfo
   {journal} {Physical Review A}\ }\textbf {\bibinfo {volume} {101}},\ \bibinfo
  {pages} {032308} (\bibinfo {year} {2020})}\BibitemShut {NoStop}%
\bibitem [{\citenamefont {Huggins}\ \emph {et~al.}(2019)\citenamefont
  {Huggins}, \citenamefont {Patil}, \citenamefont {Mitchell}, \citenamefont
  {Whaley},\ and\ \citenamefont {Stoudenmire}}]{huggins_towards_2019}%
  \BibitemOpen
  \bibfield  {author} {\bibinfo {author} {\bibfnamefont {W.}~\bibnamefont
  {Huggins}}, \bibinfo {author} {\bibfnamefont {P.}~\bibnamefont {Patil}},
  \bibinfo {author} {\bibfnamefont {B.}~\bibnamefont {Mitchell}}, \bibinfo
  {author} {\bibfnamefont {K.~B.}\ \bibnamefont {Whaley}}, \ and\ \bibinfo
  {author} {\bibfnamefont {E.~M.}\ \bibnamefont {Stoudenmire}},\ }\bibfield
  {title} {\enquote {\bibinfo {title} {Towards quantum machine learning with
  tensor networks},}\ }\href {\doibase 10.1088/2058-9565/aaea94} {\bibfield
  {journal} {\bibinfo  {journal} {Quantum Science and Technology}\ }\textbf
  {\bibinfo {volume} {4}},\ \bibinfo {pages} {024001} (\bibinfo {year}
  {2019})}\BibitemShut {NoStop}%
\bibitem [{\citenamefont {Uvarov}, \citenamefont {Kardashin},\ and\
  \citenamefont {Biamonte}(2019)}]{uvarov_machine_2019}%
  \BibitemOpen
  \bibfield  {author} {\bibinfo {author} {\bibfnamefont {A.}~\bibnamefont
  {Uvarov}}, \bibinfo {author} {\bibfnamefont {A.}~\bibnamefont {Kardashin}}, \
  and\ \bibinfo {author} {\bibfnamefont {J.}~\bibnamefont {Biamonte}},\
  }\bibfield  {title} {\enquote {\bibinfo {title} {Machine {Learning} {Phase}
  {Transitions} with a {Quantum} {Processor}},}\ }\href
  {http://arxiv.org/abs/1906.10155} {\bibfield  {journal} {\bibinfo  {journal}
  {arXiv:1906.10155 [cond-mat, physics:quant-ph]}\ } (\bibinfo {year}
  {2019})},\ \bibinfo {note} {arXiv: 1906.10155}\BibitemShut {NoStop}%
\bibitem [{\citenamefont {Brand{\~a}o}, \citenamefont {Harrow},\ and\
  \citenamefont {Horodecki}(2016)}]{brandao_local_2016}%
  \BibitemOpen
  \bibfield  {author} {\bibinfo {author} {\bibfnamefont {F.~G. S.~L.}\
  \bibnamefont {Brand{\~a}o}}, \bibinfo {author} {\bibfnamefont {A.~W.}\
  \bibnamefont {Harrow}}, \ and\ \bibinfo {author} {\bibfnamefont
  {M.}~\bibnamefont {Horodecki}},\ }\bibfield  {title} {\enquote {\bibinfo
  {title} {Local {Random} {Quantum} {Circuits} are {Approximate}
  {Polynomial}-{Designs}},}\ }\href {\doibase 10.1007/s00220-016-2706-8}
  {\bibfield  {journal} {\bibinfo  {journal} {Communications in Mathematical
  Physics}\ }\textbf {\bibinfo {volume} {346}},\ \bibinfo {pages} {397--434}
  (\bibinfo {year} {2016})}\BibitemShut {NoStop}%
\bibitem [{\citenamefont {Cerezo}\ \emph {et~al.}(2020)\citenamefont {Cerezo},
  \citenamefont {Sone}, \citenamefont {Volkoff}, \citenamefont {Cincio},\ and\
  \citenamefont {Coles}}]{cerezo_cost-function-dependent_2020}%
  \BibitemOpen
  \bibfield  {author} {\bibinfo {author} {\bibfnamefont {M.}~\bibnamefont
  {Cerezo}}, \bibinfo {author} {\bibfnamefont {A.}~\bibnamefont {Sone}},
  \bibinfo {author} {\bibfnamefont {T.}~\bibnamefont {Volkoff}}, \bibinfo
  {author} {\bibfnamefont {L.}~\bibnamefont {Cincio}}, \ and\ \bibinfo {author}
  {\bibfnamefont {P.~J.}\ \bibnamefont {Coles}},\ }\bibfield  {title} {\enquote
  {\bibinfo {title} {Cost-{Function}-{Dependent} {Barren} {Plateaus} in
  {Shallow} {Quantum} {Neural} {Networks}},}\ }\href
  {http://arxiv.org/abs/2001.00550} {\bibfield  {journal} {\bibinfo  {journal}
  {arXiv:2001.00550 [quant-ph]}\ } (\bibinfo {year} {2020})},\ \bibinfo {note}
  {arXiv: 2001.00550}\BibitemShut {NoStop}%
\bibitem [{\citenamefont {Nakaji}\ and\ \citenamefont
  {Yamamoto}(2020)}]{Nakaji_Yamamoto_2020}%
  \BibitemOpen
  \bibfield  {author} {\bibinfo {author} {\bibfnamefont {K.}~\bibnamefont
  {Nakaji}}\ and\ \bibinfo {author} {\bibfnamefont {N.}~\bibnamefont
  {Yamamoto}},\ }\bibfield  {title} {\enquote {\bibinfo {title} {Expressibility
  of the alternating layered ansatz for quantum computation},}\ }\href
  {http://arxiv.org/abs/2005.12537} {\bibfield  {journal} {\bibinfo  {journal}
  {arXiv:2005.12537 [quant-ph]}\ } (\bibinfo {year} {2020})},\ \bibinfo {note}
  {arXiv: 2005.12537}\BibitemShut {NoStop}%
\bibitem [{\citenamefont {Li}, \citenamefont {Chen},\ and\ \citenamefont
  {Fisher}(2019)}]{Li_Chen_Fisher_2019}%
  \BibitemOpen
  \bibfield  {author} {\bibinfo {author} {\bibfnamefont {Y.}~\bibnamefont
  {Li}}, \bibinfo {author} {\bibfnamefont {X.}~\bibnamefont {Chen}}, \ and\
  \bibinfo {author} {\bibfnamefont {M.~P.~A.}\ \bibnamefont {Fisher}},\
  }\bibfield  {title} {\enquote {\bibinfo {title} {Measurement-driven
  entanglement transition in hybrid quantum circuits},}\ }\href {\doibase
  10.1103/PhysRevB.100.134306} {\bibfield  {journal} {\bibinfo  {journal}
  {Physical Review B}\ }\textbf {\bibinfo {volume} {100}},\ \bibinfo {pages}
  {134306} (\bibinfo {year} {2019})}\BibitemShut {NoStop}%
\bibitem [{\citenamefont {Wecker}, \citenamefont {Hastings},\ and\
  \citenamefont {Troyer}(2015)}]{wecker_progress_2015}%
  \BibitemOpen
  \bibfield  {author} {\bibinfo {author} {\bibfnamefont {D.}~\bibnamefont
  {Wecker}}, \bibinfo {author} {\bibfnamefont {M.~B.}\ \bibnamefont
  {Hastings}}, \ and\ \bibinfo {author} {\bibfnamefont {M.}~\bibnamefont
  {Troyer}},\ }\bibfield  {title} {\enquote {\bibinfo {title} {Progress towards
  practical quantum variational algorithms},}\ }\href {\doibase
  10.1103/PhysRevA.92.042303} {\bibfield  {journal} {\bibinfo  {journal}
  {Physical Review A}\ }\textbf {\bibinfo {volume} {92}},\ \bibinfo {pages}
  {042303} (\bibinfo {year} {2015})}\BibitemShut {NoStop}%
\bibitem [{\citenamefont {Romero}\ \emph {et~al.}(2017)\citenamefont {Romero},
  \citenamefont {Babbush}, \citenamefont {McClean}, \citenamefont {Hempel},
  \citenamefont {Love},\ and\ \citenamefont
  {Aspuru-Guzik}}]{romero_strategies_2017}%
  \BibitemOpen
  \bibfield  {author} {\bibinfo {author} {\bibfnamefont {J.}~\bibnamefont
  {Romero}}, \bibinfo {author} {\bibfnamefont {R.}~\bibnamefont {Babbush}},
  \bibinfo {author} {\bibfnamefont {J.~R.}\ \bibnamefont {McClean}}, \bibinfo
  {author} {\bibfnamefont {C.}~\bibnamefont {Hempel}}, \bibinfo {author}
  {\bibfnamefont {P.}~\bibnamefont {Love}}, \ and\ \bibinfo {author}
  {\bibfnamefont {A.}~\bibnamefont {Aspuru-Guzik}},\ }\bibfield  {title}
  {\enquote {\bibinfo {title} {Strategies for quantum computing molecular
  energies using the unitary coupled cluster ansatz},}\ }\href
  {http://arxiv.org/abs/1701.02691} {\bibfield  {journal} {\bibinfo  {journal}
  {arXiv:1701.02691 [quant-ph]}\ } (\bibinfo {year} {2017})},\ \bibinfo {note}
  {arXiv: 1701.02691}\BibitemShut {NoStop}%
\bibitem [{\citenamefont {Taube}\ and\ \citenamefont
  {Bartlett}(2006)}]{taube_new_2006}%
  \BibitemOpen
  \bibfield  {author} {\bibinfo {author} {\bibfnamefont {A.~G.}\ \bibnamefont
  {Taube}}\ and\ \bibinfo {author} {\bibfnamefont {R.~J.}\ \bibnamefont
  {Bartlett}},\ }\bibfield  {title} {\enquote {\bibinfo {title} {New
  perspectives on unitary coupled-cluster theory},}\ }\href {\doibase
  10.1002/qua.21198} {\bibfield  {journal} {\bibinfo  {journal} {International
  Journal of Quantum Chemistry}\ }\textbf {\bibinfo {volume} {106}},\ \bibinfo
  {pages} {3393--3401} (\bibinfo {year} {2006})}\BibitemShut {NoStop}%
\bibitem [{\citenamefont {Lloyd}(2018)}]{lloyd_quantum_2018}%
  \BibitemOpen
  \bibfield  {author} {\bibinfo {author} {\bibfnamefont {S.}~\bibnamefont
  {Lloyd}},\ }\bibfield  {title} {\enquote {\bibinfo {title} {Quantum
  approximate optimization is computationally universal},}\ }\href
  {http://arxiv.org/abs/1812.11075} {\bibfield  {journal} {\bibinfo  {journal}
  {arXiv:1812.11075 [quant-ph]}\ } (\bibinfo {year} {2018})},\ \bibinfo {note}
  {arXiv: 1812.11075}\BibitemShut {NoStop}%
\bibitem [{\citenamefont {Morales}, \citenamefont {Biamonte},\ and\
  \citenamefont {Zimbor{\'a}s}(2019)}]{morales_universality_2019}%
  \BibitemOpen
  \bibfield  {author} {\bibinfo {author} {\bibfnamefont {M.~E.~S.}\
  \bibnamefont {Morales}}, \bibinfo {author} {\bibfnamefont {J.}~\bibnamefont
  {Biamonte}}, \ and\ \bibinfo {author} {\bibfnamefont {Z.}~\bibnamefont
  {Zimbor{\'a}s}},\ }\bibfield  {title} {\enquote {\bibinfo {title} {On the
  {Universality} of the {Quantum} {Approximate} {Optimization} {Algorithm}},}\
  }\href {http://arxiv.org/abs/1909.03123} {\bibfield  {journal} {\bibinfo
  {journal} {arXiv:1909.03123 [math-ph, physics:quant-ph]}\ } (\bibinfo {year}
  {2019})},\ \bibinfo {note} {arXiv: 1909.03123}\BibitemShut {NoStop}%
\bibitem [{\citenamefont {Biamonte}(2019)}]{biamonte_universal_2019}%
  \BibitemOpen
  \bibfield  {author} {\bibinfo {author} {\bibfnamefont {J.}~\bibnamefont
  {Biamonte}},\ }\bibfield  {title} {\enquote {\bibinfo {title} {Universal
  {Variational} {Quantum} {Computation}},}\ }\href
  {http://arxiv.org/abs/1903.04500} {\bibfield  {journal} {\bibinfo  {journal}
  {arXiv:1903.04500 [quant-ph]}\ } (\bibinfo {year} {2019})},\ \bibinfo {note}
  {arXiv: 1903.04500}\BibitemShut {NoStop}%
\bibitem [{\citenamefont {Kokail}\ \emph {et~al.}(2019)\citenamefont {Kokail},
  \citenamefont {Maier}, \citenamefont {van Bijnen}, \citenamefont {Brydges},
  \citenamefont {Joshi}, \citenamefont {Jurcevic}, \citenamefont {Muschik},
  \citenamefont {Silvi}, \citenamefont {Blatt}, \citenamefont {Roos},\ and\
  \citenamefont {Zoller}}]{kokail_self-verifying_2019}%
  \BibitemOpen
  \bibfield  {author} {\bibinfo {author} {\bibfnamefont {C.}~\bibnamefont
  {Kokail}}, \bibinfo {author} {\bibfnamefont {C.}~\bibnamefont {Maier}},
  \bibinfo {author} {\bibfnamefont {R.}~\bibnamefont {van Bijnen}}, \bibinfo
  {author} {\bibfnamefont {T.}~\bibnamefont {Brydges}}, \bibinfo {author}
  {\bibfnamefont {M.~K.}\ \bibnamefont {Joshi}}, \bibinfo {author}
  {\bibfnamefont {P.}~\bibnamefont {Jurcevic}}, \bibinfo {author}
  {\bibfnamefont {C.~A.}\ \bibnamefont {Muschik}}, \bibinfo {author}
  {\bibfnamefont {P.}~\bibnamefont {Silvi}}, \bibinfo {author} {\bibfnamefont
  {R.}~\bibnamefont {Blatt}}, \bibinfo {author} {\bibfnamefont {C.~F.}\
  \bibnamefont {Roos}}, \ and\ \bibinfo {author} {\bibfnamefont
  {P.}~\bibnamefont {Zoller}},\ }\bibfield  {title} {\enquote {\bibinfo {title}
  {Self-verifying variational quantum simulation of lattice models},}\ }\href
  {\doibase 10.1038/s41586-019-1177-4} {\bibfield  {journal} {\bibinfo
  {journal} {Nature}\ }\textbf {\bibinfo {volume} {569}},\ \bibinfo {pages}
  {355--360} (\bibinfo {year} {2019})}\BibitemShut {NoStop}%
\bibitem [{\citenamefont {Sweke}\ \emph {et~al.}(2019)\citenamefont {Sweke},
  \citenamefont {Wilde}, \citenamefont {Meyer}, \citenamefont {Schuld},
  \citenamefont {F{\"a}hrmann}, \citenamefont {Meynard-Piganeau},\ and\
  \citenamefont {Eisert}}]{sweke_stochastic_2019}%
  \BibitemOpen
  \bibfield  {author} {\bibinfo {author} {\bibfnamefont {R.}~\bibnamefont
  {Sweke}}, \bibinfo {author} {\bibfnamefont {F.}~\bibnamefont {Wilde}},
  \bibinfo {author} {\bibfnamefont {J.}~\bibnamefont {Meyer}}, \bibinfo
  {author} {\bibfnamefont {M.}~\bibnamefont {Schuld}}, \bibinfo {author}
  {\bibfnamefont {P.~K.}\ \bibnamefont {F{\"a}hrmann}}, \bibinfo {author}
  {\bibfnamefont {B.}~\bibnamefont {Meynard-Piganeau}}, \ and\ \bibinfo
  {author} {\bibfnamefont {J.}~\bibnamefont {Eisert}},\ }\bibfield  {title}
  {\enquote {\bibinfo {title} {Stochastic gradient descent for hybrid
  quantum-classical optimization},}\ }\href {http://arxiv.org/abs/1910.01155}
  {\bibfield  {journal} {\bibinfo  {journal} {arXiv:1910.01155 [quant-ph]}\ }
  (\bibinfo {year} {2019})},\ \bibinfo {note} {arXiv: 1910.01155}\BibitemShut
  {NoStop}%
\bibitem [{\citenamefont {Mitarai}\ \emph {et~al.}(2018)\citenamefont
  {Mitarai}, \citenamefont {Negoro}, \citenamefont {Kitagawa},\ and\
  \citenamefont {Fujii}}]{mitarai_quantum_2018}%
  \BibitemOpen
  \bibfield  {author} {\bibinfo {author} {\bibfnamefont {K.}~\bibnamefont
  {Mitarai}}, \bibinfo {author} {\bibfnamefont {M.}~\bibnamefont {Negoro}},
  \bibinfo {author} {\bibfnamefont {M.}~\bibnamefont {Kitagawa}}, \ and\
  \bibinfo {author} {\bibfnamefont {K.}~\bibnamefont {Fujii}},\ }\bibfield
  {title} {\enquote {\bibinfo {title} {Quantum {Circuit} {Learning}},}\ }\href
  {\doibase 10.1103/PhysRevA.98.032309} {\bibfield  {journal} {\bibinfo
  {journal} {Physical Review A}\ }\textbf {\bibinfo {volume} {98}},\ \bibinfo
  {pages} {032309} (\bibinfo {year} {2018})},\ \bibinfo {note} {arXiv:
  1803.00745}\BibitemShut {NoStop}%
\bibitem [{\citenamefont {Schuld}\ \emph {et~al.}(2019)\citenamefont {Schuld},
  \citenamefont {Bergholm}, \citenamefont {Gogolin}, \citenamefont {Izaac},\
  and\ \citenamefont {Killoran}}]{schuld_evaluating_2019}%
  \BibitemOpen
  \bibfield  {author} {\bibinfo {author} {\bibfnamefont {M.}~\bibnamefont
  {Schuld}}, \bibinfo {author} {\bibfnamefont {V.}~\bibnamefont {Bergholm}},
  \bibinfo {author} {\bibfnamefont {C.}~\bibnamefont {Gogolin}}, \bibinfo
  {author} {\bibfnamefont {J.}~\bibnamefont {Izaac}}, \ and\ \bibinfo {author}
  {\bibfnamefont {N.}~\bibnamefont {Killoran}},\ }\bibfield  {title} {\enquote
  {\bibinfo {title} {Evaluating analytic gradients on quantum hardware},}\
  }\href {\doibase 10.1103/PhysRevA.99.032331} {\bibfield  {journal} {\bibinfo
  {journal} {Physical Review A}\ }\textbf {\bibinfo {volume} {99}},\ \bibinfo
  {pages} {032331} (\bibinfo {year} {2019})}\BibitemShut {NoStop}%
\bibitem [{\citenamefont {Harrow}\ and\ \citenamefont
  {Napp}(2019)}]{harrow_low-depth_2019}%
  \BibitemOpen
  \bibfield  {author} {\bibinfo {author} {\bibfnamefont {A.}~\bibnamefont
  {Harrow}}\ and\ \bibinfo {author} {\bibfnamefont {J.}~\bibnamefont {Napp}},\
  }\bibfield  {title} {\enquote {\bibinfo {title} {Low-depth gradient
  measurements can improve convergence in variational hybrid quantum-classical
  algorithms},}\ }\href {http://arxiv.org/abs/1901.05374} {\bibfield  {journal}
  {\bibinfo  {journal} {arXiv:1901.05374 [quant-ph]}\ } (\bibinfo {year}
  {2019})},\ \bibinfo {note} {arXiv: 1901.05374}\BibitemShut {NoStop}%
\bibitem [{\citenamefont {McClean}\ \emph {et~al.}(2018)\citenamefont
  {McClean}, \citenamefont {Boixo}, \citenamefont {Smelyanskiy}, \citenamefont
  {Babbush},\ and\ \citenamefont {Neven}}]{mcclean_barren_2018}%
  \BibitemOpen
  \bibfield  {author} {\bibinfo {author} {\bibfnamefont {J.~R.}\ \bibnamefont
  {McClean}}, \bibinfo {author} {\bibfnamefont {S.}~\bibnamefont {Boixo}},
  \bibinfo {author} {\bibfnamefont {V.~N.}\ \bibnamefont {Smelyanskiy}},
  \bibinfo {author} {\bibfnamefont {R.}~\bibnamefont {Babbush}}, \ and\
  \bibinfo {author} {\bibfnamefont {H.}~\bibnamefont {Neven}},\ }\bibfield
  {title} {\enquote {\bibinfo {title} {Barren plateaus in quantum neural
  network training landscapes},}\ }\href {\doibase 10.1038/s41467-018-07090-4}
  {\bibfield  {journal} {\bibinfo  {journal} {Nature Communications}\ }\textbf
  {\bibinfo {volume} {9}},\ \bibinfo {pages} {4812} (\bibinfo {year}
  {2018})}\BibitemShut {NoStop}%
\bibitem [{\citenamefont {Harrow}\ and\ \citenamefont
  {Mehraban}(2018)}]{harrow_approximate_2018}%
  \BibitemOpen
  \bibfield  {author} {\bibinfo {author} {\bibfnamefont {A.}~\bibnamefont
  {Harrow}}\ and\ \bibinfo {author} {\bibfnamefont {S.}~\bibnamefont
  {Mehraban}},\ }\bibfield  {title} {\enquote {\bibinfo {title} {Approximate
  unitary t-designs by short random quantum circuits using nearest-neighbor and
  long-range gates},}\ }\href {http://arxiv.org/abs/1809.06957} {\bibfield
  {journal} {\bibinfo  {journal} {arXiv:1809.06957 [quant-ph]}\ } (\bibinfo
  {year} {2018})},\ \bibinfo {note} {arXiv: 1809.06957}\BibitemShut {NoStop}%
\bibitem [{\citenamefont {Khatri}\ \emph {et~al.}(2019)\citenamefont {Khatri},
  \citenamefont {LaRose}, \citenamefont {Poremba}, \citenamefont {Cincio},
  \citenamefont {Sornborger},\ and\ \citenamefont
  {Coles}}]{khatri_quantum-assisted_2018}%
  \BibitemOpen
  \bibfield  {author} {\bibinfo {author} {\bibfnamefont {S.}~\bibnamefont
  {Khatri}}, \bibinfo {author} {\bibfnamefont {R.}~\bibnamefont {LaRose}},
  \bibinfo {author} {\bibfnamefont {A.}~\bibnamefont {Poremba}}, \bibinfo
  {author} {\bibfnamefont {L.}~\bibnamefont {Cincio}}, \bibinfo {author}
  {\bibfnamefont {A.~T.}\ \bibnamefont {Sornborger}}, \ and\ \bibinfo {author}
  {\bibfnamefont {P.~J.}\ \bibnamefont {Coles}},\ }\bibfield  {title} {\enquote
  {\bibinfo {title} {Quantum-assisted quantum compiling},}\ }\href {\doibase
  10.22331/q-2019-05-13-140} {\bibfield  {journal} {\bibinfo  {journal}
  {Quantum}\ }\textbf {\bibinfo {volume} {3}},\ \bibinfo {pages} {140}
  (\bibinfo {year} {2019})}\BibitemShut {NoStop}%
\bibitem [{\citenamefont {Sharma}\ \emph {et~al.}(2020)\citenamefont {Sharma},
  \citenamefont {Cerezo}, \citenamefont {Cincio},\ and\ \citenamefont
  {Coles}}]{Sharma_Cerezo_Cincio_Coles_2020}%
  \BibitemOpen
  \bibfield  {author} {\bibinfo {author} {\bibfnamefont {K.}~\bibnamefont
  {Sharma}}, \bibinfo {author} {\bibfnamefont {M.}~\bibnamefont {Cerezo}},
  \bibinfo {author} {\bibfnamefont {L.}~\bibnamefont {Cincio}}, \ and\ \bibinfo
  {author} {\bibfnamefont {P.~J.}\ \bibnamefont {Coles}},\ }\bibfield  {title}
  {\enquote {\bibinfo {title} {Trainability of dissipative perceptron-based
  quantum neural networks},}\ }\href {http://arxiv.org/abs/2005.12458}
  {\bibfield  {journal} {\bibinfo  {journal} {arXiv:2005.12458 [quant-ph]}\ }
  (\bibinfo {year} {2020})},\ \bibinfo {note} {arXiv: 2005.12458}\BibitemShut
  {NoStop}%
\bibitem [{\citenamefont {Watrous}(2018)}]{watrous_theory_2018}%
  \BibitemOpen
  \bibfield  {author} {\bibinfo {author} {\bibfnamefont {J.}~\bibnamefont
  {Watrous}},\ }\href@noop {} {\emph {\bibinfo {title} {The theory of quantum
  information}}}\ (\bibinfo  {publisher} {Cambridge University Press},\
  \bibinfo {address} {Cambridge, United Kingdom},\ \bibinfo {year}
  {2018})\BibitemShut {NoStop}%
\bibitem [{\citenamefont {Emerson}, \citenamefont {Livine},\ and\ \citenamefont
  {Lloyd}(2005)}]{emerson_convergence_2005}%
  \BibitemOpen
  \bibfield  {author} {\bibinfo {author} {\bibfnamefont {J.}~\bibnamefont
  {Emerson}}, \bibinfo {author} {\bibfnamefont {E.}~\bibnamefont {Livine}}, \
  and\ \bibinfo {author} {\bibfnamefont {S.}~\bibnamefont {Lloyd}},\ }\bibfield
   {title} {\enquote {\bibinfo {title} {Convergence conditions for random
  quantum circuits},}\ }\href {\doibase 10.1103/PhysRevA.72.060302} {\bibfield
  {journal} {\bibinfo  {journal} {Physical Review A}\ }\textbf {\bibinfo
  {volume} {72}},\ \bibinfo {pages} {060302} (\bibinfo {year}
  {2005})}\BibitemShut {NoStop}%
\bibitem [{\citenamefont {Collins}\ and\ \citenamefont
  {{\'S}niady}(2006)}]{collins_integration_2006}%
  \BibitemOpen
  \bibfield  {author} {\bibinfo {author} {\bibfnamefont {B.}~\bibnamefont
  {Collins}}\ and\ \bibinfo {author} {\bibfnamefont {P.}~\bibnamefont
  {{\'S}niady}},\ }\bibfield  {title} {\enquote {\bibinfo {title} {Integration
  with {Respect} to the {Haar} {Measure} on {Unitary}, {Orthogonal} and
  {Symplectic} {Group}},}\ }\href {\doibase 10.1007/s00220-006-1554-3}
  {\bibfield  {journal} {\bibinfo  {journal} {Communications in Mathematical
  Physics}\ }\textbf {\bibinfo {volume} {264}},\ \bibinfo {pages} {773--795}
  (\bibinfo {year} {2006})}\BibitemShut {NoStop}%
\bibitem [{\citenamefont {Samuel}(1980)}]{Samuel_1980}%
  \BibitemOpen
  \bibfield  {author} {\bibinfo {author} {\bibfnamefont {S.}~\bibnamefont
  {Samuel}},\ }\bibfield  {title} {\enquote {\bibinfo {title} {U( n )
  integrals, 1/ n , and the de wit–’t hooft anomalies},}\ }\href {\doibase
  10.1063/1.524386} {\bibfield  {journal} {\bibinfo  {journal} {Journal of
  Mathematical Physics}\ }\textbf {\bibinfo {volume} {21}},\ \bibinfo {pages}
  {2695–2703} (\bibinfo {year} {1980})}\BibitemShut {NoStop}%
\bibitem [{\citenamefont {Low}(2010)}]{Low_2010}%
  \BibitemOpen
  \bibfield  {author} {\bibinfo {author} {\bibfnamefont {R.~A.}\ \bibnamefont
  {Low}},\ }\bibfield  {title} {\enquote {\bibinfo {title} {Pseudo-randomness
  and learning in quantum computation},}\ }\href
  {http://arxiv.org/abs/1006.5227} {\bibfield  {journal} {\bibinfo  {journal}
  {arXiv:1006.5227 [quant-ph]}\ } (\bibinfo {year} {2010})},\ \bibinfo {note}
  {arXiv: 1006.5227}\BibitemShut {NoStop}%
\bibitem [{\citenamefont {Poland}, \citenamefont {Beer},\ and\ \citenamefont
  {Osborne}(2020)}]{Poland_Beer_Osborne_2020}%
  \BibitemOpen
  \bibfield  {author} {\bibinfo {author} {\bibfnamefont {K.}~\bibnamefont
  {Poland}}, \bibinfo {author} {\bibfnamefont {K.}~\bibnamefont {Beer}}, \ and\
  \bibinfo {author} {\bibfnamefont {T.~J.}\ \bibnamefont {Osborne}},\
  }\bibfield  {title} {\enquote {\bibinfo {title} {No free lunch for quantum
  machine learning},}\ }\href {http://arxiv.org/abs/2003.14103} {\bibfield
  {journal} {\bibinfo  {journal} {arXiv:2003.14103 [quant-ph]}\ } (\bibinfo
  {year} {2020})},\ \bibinfo {note} {arXiv: 2003.14103}\BibitemShut {NoStop}%
\bibitem [{\citenamefont {Khaneja}\ and\ \citenamefont
  {Glaser}(2000)}]{khaneja_cartan_2000}%
  \BibitemOpen
  \bibfield  {author} {\bibinfo {author} {\bibfnamefont {N.}~\bibnamefont
  {Khaneja}}\ and\ \bibinfo {author} {\bibfnamefont {S.}~\bibnamefont
  {Glaser}},\ }\bibfield  {title} {\enquote {\bibinfo {title} {Cartan
  {Decomposition} of {SU}(2{\textasciicircum}n), {Constructive}
  {Controllability} of {Spin} systems and {Universal} {Quantum} {Computing}},}\
  }\href {http://arxiv.org/abs/quant-ph/0010100} {\bibfield  {journal}
  {\bibinfo  {journal} {arXiv:quant-ph/0010100}\ } (\bibinfo {year} {2000})},\
  \bibinfo {note} {arXiv: quant-ph/0010100}\BibitemShut {NoStop}%
\bibitem [{\citenamefont {Khaneja}, \citenamefont {Brockett},\ and\
  \citenamefont {Glaser}(2001)}]{khaneja_time_2001}%
  \BibitemOpen
  \bibfield  {author} {\bibinfo {author} {\bibfnamefont {N.}~\bibnamefont
  {Khaneja}}, \bibinfo {author} {\bibfnamefont {R.}~\bibnamefont {Brockett}}, \
  and\ \bibinfo {author} {\bibfnamefont {S.~J.}\ \bibnamefont {Glaser}},\
  }\bibfield  {title} {\enquote {\bibinfo {title} {Time optimal control in spin
  systems},}\ }\href {\doibase 10.1103/PhysRevA.63.032308} {\bibfield
  {journal} {\bibinfo  {journal} {Physical Review A}\ }\textbf {\bibinfo
  {volume} {63}},\ \bibinfo {pages} {032308} (\bibinfo {year}
  {2001})}\BibitemShut {NoStop}%
\bibitem [{\citenamefont {Ryabinkin}\ \emph {et~al.}(2020)\citenamefont
  {Ryabinkin}, \citenamefont {Lang}, \citenamefont {Genin},\ and\ \citenamefont
  {Izmaylov}}]{ryabinkin_iterative_2019}%
  \BibitemOpen
  \bibfield  {author} {\bibinfo {author} {\bibfnamefont {I.~G.}\ \bibnamefont
  {Ryabinkin}}, \bibinfo {author} {\bibfnamefont {R.~A.}\ \bibnamefont {Lang}},
  \bibinfo {author} {\bibfnamefont {S.~N.}\ \bibnamefont {Genin}}, \ and\
  \bibinfo {author} {\bibfnamefont {A.~F.}\ \bibnamefont {Izmaylov}},\
  }\bibfield  {title} {\enquote {\bibinfo {title} {Iterative qubit coupled
  cluster approach with efficient screening of generators},}\ }\href {\doibase
  10.1021/acs.jctc.9b01084} {\bibfield  {journal} {\bibinfo  {journal} {Journal
  of Chemical Theory and Computation}\ }\textbf {\bibinfo {volume} {16}},\
  \bibinfo {pages} {1055–1063} (\bibinfo {year} {2020})}\BibitemShut
  {NoStop}%
\bibitem [{\citenamefont {Verteletskyi}, \citenamefont {Yen},\ and\
  \citenamefont {Izmaylov}(2020)}]{verteletskyi_measurement_2019}%
  \BibitemOpen
  \bibfield  {author} {\bibinfo {author} {\bibfnamefont {V.}~\bibnamefont
  {Verteletskyi}}, \bibinfo {author} {\bibfnamefont {T.-C.}\ \bibnamefont
  {Yen}}, \ and\ \bibinfo {author} {\bibfnamefont {A.~F.}\ \bibnamefont
  {Izmaylov}},\ }\bibfield  {title} {\enquote {\bibinfo {title} {Measurement
  optimization in the variational quantum eigensolver using a minimum clique
  cover},}\ }\href {\doibase 10.1063/1.5141458} {\bibfield  {journal} {\bibinfo
   {journal} {The Journal of Chemical Physics}\ }\textbf {\bibinfo {volume}
  {152}},\ \bibinfo {pages} {124114} (\bibinfo {year} {2020})}\BibitemShut
  {NoStop}%
\bibitem [{\citenamefont {Grant}\ \emph {et~al.}(2019)\citenamefont {Grant},
  \citenamefont {Wossnig}, \citenamefont {Ostaszewski},\ and\ \citenamefont
  {Benedetti}}]{grant_initialization_2019}%
  \BibitemOpen
  \bibfield  {author} {\bibinfo {author} {\bibfnamefont {E.}~\bibnamefont
  {Grant}}, \bibinfo {author} {\bibfnamefont {L.}~\bibnamefont {Wossnig}},
  \bibinfo {author} {\bibfnamefont {M.}~\bibnamefont {Ostaszewski}}, \ and\
  \bibinfo {author} {\bibfnamefont {M.}~\bibnamefont {Benedetti}},\ }\bibfield
  {title} {\enquote {\bibinfo {title} {An initialization strategy for
  addressing barren plateaus in parametrized quantum circuits},}\ }\href
  {\doibase 10.22331/q-2019-12-09-214} {\bibfield  {journal} {\bibinfo
  {journal} {Quantum}\ }\textbf {\bibinfo {volume} {3}},\ \bibinfo {pages}
  {214} (\bibinfo {year} {2019})},\ \bibinfo {note} {arXiv:
  1903.05076}\BibitemShut {NoStop}%
\bibitem [{\citenamefont {Skolik}\ \emph {et~al.}(2020)\citenamefont {Skolik},
  \citenamefont {McClean}, \citenamefont {Mohseni}, \citenamefont {van~der
  Smagt},\ and\ \citenamefont {Leib}}]{Skolik_2020}%
  \BibitemOpen
  \bibfield  {author} {\bibinfo {author} {\bibfnamefont {A.}~\bibnamefont
  {Skolik}}, \bibinfo {author} {\bibfnamefont {J.~R.}\ \bibnamefont {McClean}},
  \bibinfo {author} {\bibfnamefont {M.}~\bibnamefont {Mohseni}}, \bibinfo
  {author} {\bibfnamefont {P.}~\bibnamefont {van~der Smagt}}, \ and\ \bibinfo
  {author} {\bibfnamefont {M.}~\bibnamefont {Leib}},\ }\bibfield  {title}
  {\enquote {\bibinfo {title} {Layerwise learning for quantum neural
  networks},}\ }\href {http://arxiv.org/abs/2006.14904} {\bibfield  {journal}
  {\bibinfo  {journal} {arXiv:2006.14904 [quant-ph]}\ } (\bibinfo {year}
  {2020})},\ \bibinfo {note} {arXiv: 2006.14904}\BibitemShut {NoStop}%
\bibitem [{\citenamefont {Garcia-Saez}\ and\ \citenamefont
  {Latorre}(2018)}]{garcia-saez_addressing_2018}%
  \BibitemOpen
  \bibfield  {author} {\bibinfo {author} {\bibfnamefont {A.}~\bibnamefont
  {Garcia-Saez}}\ and\ \bibinfo {author} {\bibfnamefont {J.~I.}\ \bibnamefont
  {Latorre}},\ }\bibfield  {title} {\enquote {\bibinfo {title} {Addressing hard
  classical problems with {Adiabatically} {Assisted} {Variational} {Quantum}
  {Eigensolvers}},}\ }\href {http://arxiv.org/abs/1806.02287} {\bibfield
  {journal} {\bibinfo  {journal} {arXiv:1806.02287 [cond-mat,
  physics:quant-ph]}\ } (\bibinfo {year} {2018})},\ \bibinfo {note} {arXiv:
  1806.02287}\BibitemShut {NoStop}%
\bibitem [{\citenamefont {Bravo-Prieto}\ \emph {et~al.}(2020)\citenamefont
  {Bravo-Prieto}, \citenamefont {Lumbreras-Zarapico}, \citenamefont
  {Tagliacozzo},\ and\ \citenamefont {Latorre}}]{bravo-prieto_scaling_2020}%
  \BibitemOpen
  \bibfield  {author} {\bibinfo {author} {\bibfnamefont {C.}~\bibnamefont
  {Bravo-Prieto}}, \bibinfo {author} {\bibfnamefont {J.}~\bibnamefont
  {Lumbreras-Zarapico}}, \bibinfo {author} {\bibfnamefont {L.}~\bibnamefont
  {Tagliacozzo}}, \ and\ \bibinfo {author} {\bibfnamefont {J.~I.}\ \bibnamefont
  {Latorre}},\ }\bibfield  {title} {\enquote {\bibinfo {title} {Scaling of
  variational quantum circuit depth for condensed matter systems},}\ }\href
  {http://arxiv.org/abs/2002.06210} {\bibfield  {journal} {\bibinfo  {journal}
  {arXiv:2002.06210 [cond-mat, physics:quant-ph]}\ } (\bibinfo {year}
  {2020})},\ \bibinfo {note} {arXiv: 2002.06210}\BibitemShut {NoStop}%
\bibitem [{\citenamefont {Bilkis}\ \emph {et~al.}(2021)\citenamefont {Bilkis},
  \citenamefont {Cerezo}, \citenamefont {Verdon}, \citenamefont {Coles},\ and\
  \citenamefont {Cincio}}]{Bilkis_Cerezo_Verdon_Coles_Cincio_2021}%
  \BibitemOpen
  \bibfield  {author} {\bibinfo {author} {\bibfnamefont {M.}~\bibnamefont
  {Bilkis}}, \bibinfo {author} {\bibfnamefont {M.}~\bibnamefont {Cerezo}},
  \bibinfo {author} {\bibfnamefont {G.}~\bibnamefont {Verdon}}, \bibinfo
  {author} {\bibfnamefont {P.~J.}\ \bibnamefont {Coles}}, \ and\ \bibinfo
  {author} {\bibfnamefont {L.}~\bibnamefont {Cincio}},\ }\bibfield  {title}
  {\enquote {\bibinfo {title} {A semi-agnostic ansatz with variable structure
  for quantum machine learning},}\ }\href {http://arxiv.org/abs/2103.06712}
  {\bibfield  {journal} {\bibinfo  {journal} {arXiv:2103.06712 [quant-ph,
  stat]}\ } (\bibinfo {year} {2021})},\ \bibinfo {note} {arXiv:
  2103.06712}\BibitemShut {NoStop}%
\bibitem [{\citenamefont {Grimsley}\ \emph {et~al.}(2019)\citenamefont
  {Grimsley}, \citenamefont {Economou}, \citenamefont {Barnes},\ and\
  \citenamefont {Mayhall}}]{grimsley_adaptive_2019}%
  \BibitemOpen
  \bibfield  {author} {\bibinfo {author} {\bibfnamefont {H.~R.}\ \bibnamefont
  {Grimsley}}, \bibinfo {author} {\bibfnamefont {S.~E.}\ \bibnamefont
  {Economou}}, \bibinfo {author} {\bibfnamefont {E.}~\bibnamefont {Barnes}}, \
  and\ \bibinfo {author} {\bibfnamefont {N.~J.}\ \bibnamefont {Mayhall}},\
  }\bibfield  {title} {\enquote {\bibinfo {title} {An adaptive variational
  algorithm for exact molecular simulations on a quantum computer},}\ }\href
  {\doibase 10.1038/s41467-019-10988-2} {\bibfield  {journal} {\bibinfo
  {journal} {Nature Communications}\ }\textbf {\bibinfo {volume} {10}},\
  \bibinfo {pages} {3007} (\bibinfo {year} {2019})}\BibitemShut {NoStop}%
\end{thebibliography}%

\newpage

\appendix

\section{Meaning of the operator 2-norm of the TPE}
\label{sec:2-norm_TPE}

Let $H = \sum c_i \sigma_i$ be a Hamiltonian on $n$ qubits. Recall that Pauli strings form an orthogonal basis. Since the Hilbert-Schmidt inner product is the same as the scalar product of matrices as vectors in $\mathbb{R}^{2^n \times 2^n}$, this also applies to their reshaping to vectors. Then one can verify that $||\mathrm{vec}(H)||_2 = 
2^{\frac{n}{2}} \sqrt{\sum_i |c_i|^2}
\equiv ||\mathbf{c}||_2 \cdot 2^{\frac{n}{2}}$. This works when $\sigma_i$ are super Pauli strings. The operator 2-norm of $\mathbb{E}_{Haar} (U^{\otimes t} \otimes (U^*)^{\otimes t}) - \mathbb{E}_{\mu} (U^{\otimes t} \otimes (U^*)^{\otimes t})$, provides an upper bound on the vector norm of the output of this operator, meaning that this is the maximum norm of the discrepancy from the perfect output for an input of unit norm. For a super Pauli string $h \otimes h$, this error $\lambda_2$ upper bounds the 2-norm of the vector $(\mathbf{c} - \mathbf{c}_{Haar})$.

\section{Structure of the two-dimensional lattice ansatz}
\label{sec:2d_ansatz}

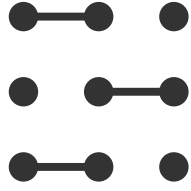
\begin{figure}
    \centering
    \begin{tikzpicture}[thick]
    \node at ( 0,0) [circle,draw=black!80,fill=black!80] {};
    \node at ( 0,1) [circle,draw=black!80,fill=black!80] {};
    \node at ( 0,2) [circle,draw=black!80,fill=black!80] {};
    \node at ( 1,0) [circle,draw=black!80,fill=black!80] {};
    \node at ( 1,1) [circle,draw=black!80,fill=black!80] {};
    \node at ( 1,2) [circle,draw=black!80,fill=black!80] {};
    \node at ( 2,0) [circle,draw=black!80,fill=black!80] {};
    \node at ( 2,1) [circle,draw=black!80,fill=black!80] {};
    \node at ( 2,2) [circle,draw=black!80,fill=black!80] {};
    
    \draw [draw=black!80,line width=3] (0,0) -- (1,0);
    \draw [draw=black!80,line width=3] (1,1) -- (2,1);
    \draw [draw=black!80,line width=3] (0,2) -- (1,2);
    \end{tikzpicture}    
    \caption{Connectivity of one layer in the 2D lattice ansatz. Other layers are formed by rotating this pattern by 90 degrees.}
    \label{fig:2d_ansatz_scheme}
\end{figure}

The two-dimensional ansatz consisted of four layers of two-qubit blocks. The first layer of the ansatz is schematically depicted in Fig.~\ref{fig:2d_ansatz_scheme}. Every next layer is obtained from the last by rotating the layout 90 degrees clockwise.

\end{document}